\documentclass[fleqn,usenatbib]{mnras}

\usepackage{newtxtext,newtxmath}

\usepackage[T1]{fontenc}

\DeclareRobustCommand{\VAN}[3]{#2}
\let\VANthebibliography\thebibliography
\def\thebibliography{\DeclareRobustCommand{\VAN}[3]{##3}\VANthebibliography}

\usepackage{graphicx}	
\usepackage{amsmath}	

\usepackage{bm}
\usepackage{float}

\usepackage{xcolor}

\newcommand{\specialcell}[2][c]{%
  \begin{tabular}[#1]{@{}c@{}}#2\end{tabular}}

\title[Multi-Class Anomaly Detection]{A Classifier-Based Approach to Multi-Class Anomaly Detection for Astronomical Transients}

\author[R. Gupta et al.]{
Rithwik Gupta,$^{2}$
Daniel Muthukrishna,$^{1}$\thanks{E-mail: \href{mailto:danmuth@mit.edu}{danmuth@mit.edu}}
and Michelle Lochner$^{3,4}$
\\
$^{1}$Kavli Institute for Astrophysics and Space Research, Massachusetts Institute of Technology, Cambridge, MA 02139, USA\\
$^{2}$Irvington High School, 41800 Blacow Rd, Fremont, CA 94538, USA\\
$^{3}$Department of Physics and Astronomy, University of the Western Cape, Bellville, Cape Town, 7535, South Africa\\
$^{4}$South African Radio Astronomy Observatory, 2 Fir Street, Black River Park, Observatory, 7925, South Africa\\
}

\date{Accepted 18/11/2024. Received 04/09/2024 ; in original form 22/04/2024}

\pubyear{2024}

\begin{document}
\label{firstpage}
\pagerange{\pageref{firstpage}--\pageref{lastpage}}
\maketitle

\graphicspath{ {./images/} }

\begin{abstract}
Automating real-time anomaly detection is essential for identifying rare transients, with modern survey telescopes generating tens of thousands of alerts per night, and future telescopes, such as the Vera C. Rubin Observatory, projected to increase this number dramatically. Currently, most anomaly detection algorithms for astronomical transients rely either on hand-crafted features extracted from light curves or on features generated through unsupervised representation learning, coupled with standard anomaly detection algorithms. In this work, we introduce an alternative approach: using the penultimate layer of a neural network classifier as the latent space for anomaly detection. We then propose a novel method, Multi-Class Isolation Forests (\texttt{MCIF}), which trains separate isolation forests for each class to derive an anomaly score for a light curve from its latent space representation. This approach significantly outperforms a standard isolation forest. We also use a simpler input method for real-time transient classifiers which circumvents the need for interpolation and helps the neural network handle irregular sampling and model inter-passband relationships. Our anomaly detection pipeline identifies rare classes including kilonovae, pair-instability supernovae, and intermediate luminosity transients shortly after trigger on simulated Zwicky Transient Facility light curves. Using a sample of our simulations matching the population of anomalies expected in nature (54 anomalies and 12,040 common transients), our method discovered $41\pm3$ anomalies ($\sim75\%$ recall) after following up the top 2000 ($\sim15\%$) ranked transients. Our novel method shows that classifiers can be effectively repurposed for real-time anomaly detection.

\end{abstract}

\begin{keywords}
Machine Learning -- Data Methods -- Software -- Transients -- Supernovae -- Time-Domain Astronomy
\end{keywords}



\section{Introduction}
\label{sec:introduction}

With the advancement of survey telescopes and the advent of large-scale transient surveys, we are entering a new paradigm for astronomical study. The Vera Rubin Observatory's Legacy Survey of Space and Time (LSST) is expected to generate ten million transient alerts per night \citep{Ivezic2009LSST:Products}. The traditional approach of manual examination of astronomical data, which has led to some of the biggest discoveries in astronomy, is no longer feasible. As a result, there is a growing need to develop methods that can automate the serendipity that has so far played a pivotal role in scientific discovery. Furthermore, in the case of transients, it is also imperative that anomalies can be identified in real time so that new astrophysical phenomena can be discovered early and studied at each stage of their evolution. In particular, detailed early-time follow-up is necessary to understand the progenitor systems and explosion mechanisms of transients \citep[e.g.][]{Kasen2010}. Currently, the central engines of many rare transient classes, such as calcium-rich transients, kilonovae, and the newly discovered fast blue optical transients (FBOTs) \citep[e.g.][]{CoppejansMargutti2020FBOTs} remain poorly understood. This, along with the considerable amount of human effort in the follow-up of GW170817 \citep[e.g.][]{2017ApJ...848L..12A}, the first observed binary neutron star merger, has reinforced the need for automatic photometric identification of new astrophysical phenomena.

The announcement of LSST motivated the development of many real-time \citep[e.g.][]{Narayan2018MachineStream, SupernnoovaMoller2019, Muthukrishna19RAPID} and full light-curve classifiers \citep[e.g.][]{FullLightCurveClass1, PELICANPasquet2019, Lochner2016}. An inherent quality of classifiers is the fundamental assumption that all observed objects belong to one of the predefined classes; however, this is not the case. New telescopes are finding interesting new types of astronomical objects by probing deeper, wider, and faster than ever before. For example, LSST will have an unprecedented point-source depth of $r \sim 27.5$ \citep{Ivezic2009LSST:Products}, probing fainter than any other survey to date, and the Transiting Exoplanet Survey Satellite \citep[TESS][]{TESS_Ricker_2015} is exploring transients at a much shorter time scale, from hours to minutes, using its wide field-of-view. Astronomers will need automatic tools to assist in identifying which potentially anomalous events to follow up.

In this regard, anomaly detection is most simply defined as identifying outlier samples. While this may be straightforward in low-dimensional spaces, it becomes considerably more challenging when dealing with multi-passband astronomical light curves, which typically feature a large and often variable number of inputs. Thus, most previous studies in anomaly detection attempt to find a lower-dimensional latent space that is easier to cluster and identify anomalies. Previous works often use either user-defined feature extraction \citep[e.g.][]{Webb2020,Giles2019_Timeseries, Ishida2021_Timeseries, Malanchev2021, Pruzhinskaya2019, AnomFullCurve1, Perez-Carrasco_2023} or deep learning \citep{vraenn, Malanchev2021, Solarz2017} to encode this latent space. In this work, we employ deep learning for feature extraction, which is quickly becoming the gold standard in the field.  Deep learning is emerging as the preferred approach due to its ability to automatically learn complex, hierarchical features from raw data without the need for manual feature engineering. This is particularly advantageous in astronomy, where the underlying physical processes generating the data are often not fully understood, making it difficult to design comprehensive user-defined features. These data-driven models can also adapt to new data distributions and scale efficiently with increasing data volumes allowing for more generalizable models that can be applied across various astronomical objects and observational scenarios with minimal modification.

Throughout the literature on anomaly detection for astronomical transients, two different definitions of the problem are presented. Some approaches, categorized as unsupervised methods, focus on extracting anomalies from large datasets without relying on prior information \citep[e.g.][]{vraenn, Webb2020, Giles2019_Timeseries}. Numerous differing approaches exist for unsupervised anomaly detection. \cite{vraenn} used an unsupervised recurrent variational autoencoder to find a representative latent space mapping of the light curves to then derive anomaly scores using an isolation forest. \cite{Webb2020} used user-defined feature extraction and then an isolation forest with active learning to find anomalies.

In contrast, our work, among others \cite[e.g.][]{Perez-Carrasco_2023, OriginalPaper}, utilizes previous, either simulated or real, transients to determine whether a new light curve is anomalous. This approach is often referred to as novelty detection or supervised anomaly detection. Previous novelty detection approaches \citep[e.g.][]{OriginalPaper, Soraism2020Novelties} are often variations of one-class classification \citep{OneClassDef}. One-class classifiers attempt to model a set of \textit{normal} samples and then classify new transients as either part of that sample or as outliers. One-class methods have been shown to be effective at anomaly detection \citep{ruff_2018_OneClass}, but they do not capture the complexity of the population of known astronomical transients, that are grouped into numerous classes with intrinsically different qualities. It is challenging for an algorithm to classify this diverse population of known transients into a single class and still identify anomalies. \cite{Perez-Carrasco_2023} released a method to combat this issue. It extended the one-class classifier to multiple classes on features extracted from full multi-passband light curve data. Their method adapts the single-class loss function to multiple classes by encouraging light curves of the same class to cluster together. The loss function of their encoder becomes a measure of how clustered objects of the same class are in the latent space, and extracts anomalies by using the minimum distance to any cluster in the latent space.

The announcement of LSST has also made real-time anomaly detection more important \citep[e.g.][]{vraenn, OriginalPaper, Soraism2020Novelties, AnomRealTimeGWAC1, AnomRealTimeGWAC2}. \cite{vraenn} generalized their variational autoencoder to use a recurrent neural network, allowing real-time anomaly scores. \cite{OriginalPaper} used predictive modeling and derived the anomaly score as the deviation from real-time predictions. \cite{Soraism2020Novelties} used magnitude changes in real time to assess the probability of a new transient being similar to the common transient sample.

In this work, we leverage a light-curve classifier to address the one-class challenge and distinguish between the various classes of transients. Our approach demonstrates promising clustering in the feature space, the penultimate layer of the classifier, and shows a substantial level of discrimination in anomaly scores. Notably, similar feature extraction methods have shown potential in the field of astronomical image analysis \citep[e.g.][]{etsebeth2023astronomaly, WalmsleyCNN}.

Previous light-curve classification works exist in both the domain of real-time classification \citep[e.g.][]{Muthukrishna19RAPID, SupernnoovaMoller2019, Mahabal2008} and full light-curve classification \citep[e.g.][]{Charnock2016, PELICANPasquet2019}. Real-time classifiers predict a classification output at every new observation, while full light-curve classifiers retrospectively predict a classification output on a full light-curve. Most previous real-time approaches employ some interpolation technique which serves as the bottleneck for the model. Our real-time classifier, on the other hand, uses a novel input method in this domain, which omits the use of interpolation and can help the neural network understand the relationship between different passbands.

After identifying a feature space using one of the aforementioned methods, several prior works have employed an isolation forest \citep{isolationforest} to generate anomaly scores. Isolation forests are one of the most popular anomaly detection architectures. They work by recursively partitioning data using random splits, with the idea that outliers are rare and different, thus requiring fewer partitions to isolate them. The algorithm assigns an anomaly score based on the average path length needed to isolate each object, with shorter paths indicating potential anomalies.

While Isolation Forests have demonstrated success in previous research within the same domain \citep[e.g][]{vraenn, Ishida2021_Timeseries, IsoForestUse2}, it, too, faces challenges when dealing with a complex latent space housing multiple clusters of intrinsically different transient classes. Consequently, the application of a single isolation forest may have limitations in accurately identifying certain anomalies. \citet{singh2022multiclass} also recognized problems in a general machine learning context, and introduced a pipeline where an autoencoder is trained as an anomaly detector on observations from every class, treating all other observations as anomalous. The final anomaly score is determined as the minimum score from all detectors. This method has shown promising results in comparison to other anomaly detection methods.

In response to this limitation and following a similar principle to \cite{singh2022multiclass}, we propose the use of Multi-Class Isolation Forests (\texttt{MCIF}): a method that involves training a separate isolation forest for each known class and extracting the minimum score among them as the final anomaly score for a given sample. Our experimental results suggest that \texttt{MCIF} can improve anomaly detection performance for certain types of optical transients.

The paper is organized as follows. In \S \ref{sec:Data}, we discuss the simulated data used and how it is preprocessed to fit real-time detection. In \S \ref{sec:Methods} we provide an overview of the proposed architecture, with \S \ref{sec:ClassifierMethod} detailing the classifier and \S \ref{sec:AnomalyMethod} introducing Multi-Class Isolation Forests (\texttt{MCIF}). Similarly, \S \ref{sec:Results} discusses the results of our approach, with \S \ref{sec:ClassifierRes} analyzing the classifier and \S \ref{sec:AnomalyRes} evaluating the anomaly detection pipeline. We conclude in \S \ref{sec:Conclusion} by outlining avenues for future work. Finally, we show the advantages of \texttt{MCIF} compared to a normal isolation forest in Appendix \ref{sec:MCIF_Advantages}.

\section{Data}

\label{sec:Data}

\subsection{Simulated Data}
In this work, we use a collection of simulated light curves that match the observing properties of the Zwicky Transient Facility \citep[ZTF;][]{ZTF}. This dataset is described in \S~2 of \citet{OriginalPaper} and is based on the simulations developed for PLAsTiCC \citep{PlasticcSim}. Each transient in the dataset has flux and flux error measurements in the $g$ and $r$ passbands with a median cadence of roughly 3 days in each passband\footnote{The public MSIP ZTF survey has since changed to a 2-day median cadence.}. We briefly describe the 17 transient classes from \citet{KesslerPlasticcModels} that are used in this work. Example light curves from each of these classes are illustrated in Appendix \ref{sec:dataset}, Figs.~1-3 of \citet{KesslerPlasticcModels}, and Fig.~2 of \citet{Muthukrishna19RAPID}.

\begin{enumerate}
    \item \textbf{Type Ia Supernovae} (SNIa) involves a carbon-oxygen white dwarf star accreting mass from a binary companion star. The white dwarf star eventually reaches a critical mass that triggers an explosion.
    \item \textbf{Type Ia-91bg Supernovae} (SNIa-91bg) tend to have fast-evolving light curves and often have lower luminosities when compared with typical SNIa.
    \item \textbf{Type Iax Supernovae} (SNIax) tend to have lower luminosity and slower ejecta velocities than typical SNIa. 
    \item \textbf{Type Ib and Ic Supernovae} (SNIb and SNIc) are thought to be caused by the core collapse of highly dense stars. Both SNIb and SNIc have lost their hydrogen envelopes prior to collapse, but SNIc have also lost their helium envelopes. These SNe are characterized by the lack of hydrogen emissions in their spectra.
    \item \textbf{Type Ic-BL Supernovae} (SNIc-BL) are SNIc with broad lines in their spectra, indicating much faster expansion velocities than typical SNIc.
    \item \textbf{Type II Supernovae} (SNII) are thought to also be formed by the core-collapses of highly dense stars. However, unlike SNIb and SNIc, SNII have hydrogen emission lines in their spectra.
    \item \textbf{Type IIb Supernovae} (SNIIb) appear very similar to SNIb. They have rapidly fading hydrogen emission lines resulting in a very similar structure to SNIb.
    \item \textbf{Type IIn Supernovae} (SNIIn) are SNII with very narrow hydrogen emission lines. 
    \item \textbf{Type I Super Luminous Supernovae} (SLSN-I) are poorly understood and very bright SNe events. SLSN-I lack hydrogen lines in their spectra.
    \item \textbf{Pair Instability Supernovae} (PISN) are the result of the explosion of a massive star, much larger than a SNII or SNIb (130 to 250 solar masses).
    \item \textbf{Kilonovae} (KN) are explosions resulting from the merging of two neutron stars or a neutron star and a black hole. Only one KN has been observed to date.
    \item \textbf{Active Galactic Nuclei} (AGN) are the very bright centers of galaxies where the supermassive black hole accretes material and emits significant radiation across the electromagnetic spectrum.
    \item \textbf{Tidal Disruption Events} (TDE) occur when a star orbiting a black hole is pulled apart by the black hole's tidal forces. The bright flare caused by this event can last up to a few years.  
    \item \textbf{Intermediate Luminosity Optical Transients} (ILOT) are a very poorly understood transient. They occur in the energy gap between normal novae and supernovae.
    \item \textbf{Calcium Rich Transients} (CaRT) are defined by their strong calcium emission lines. Their lower luminosity than SNIa and their rapidly evolving rise in luminosity after explosion make their light curves have a strong resemblance to core-collapse supernovae (SNIb and SNIc) and some SNIa-91bg.
    \item \textbf{$\mu$Lens-BSR} (uLens-BSR) are a special case of microlensing events where a binary system in the foreground acts as the lens for a background star. The light curves from these events can exhibit asymmetries, multiple peaks, plateaus, and quasiperiodic behavior.
    
\end{enumerate}

Due to their low occurrence in nature\footnote{The expected relative frequencies of each class are taken from \citealp{PlasticcSim} developed for the PLaSTiCC Dataset.}, \textbf{KN, ILOT, CaRT, PISN, and uLens-BSR} are considered the \textbf{anomalous classes} in this work, and all remaining classes are considered the ``common'' classes. However, note that the goal of this work is not to identify these specific anomalous classes, but rather identify anomalies in general, as discussed further in \S \ref{sec:curse-of-anom}.


\begin{table}
\begin{center}

\caption{Number of transients in the training set, validation set, test set, and realistic samples (see Section \ref{sec:representative_population}) for each class. All anomalous data is reserved for evaluation.}
\label{table:populationdistribution} 
\begin{tabular}{||c c c c c c||} 
 \hline
 Class & Training & Validation & Test & Total & \specialcell{Realistic\\Sample$^a$} \\ [0.5ex] 
 \hline\hline
SNIa & 9314 & 1131 & 1142 & 11587 & 1142 \\
\hline
SNIa-91bg & 10361 & 1318 & 1321 & 13000 & 1318 \\
\hline
SNIax & 10413 & 1248 & 1339 & 13000 & 1339 \\
\hline
SNIb & 4197 & 507 & 563 & 5267 & 563 \\
\hline
SNIc & 1279 & 169 & 135 & 1583 & 135 \\
\hline
SNIc-BL & 1157 & 124 & 142 & 1423 & 142 \\
\hline
SNII & 10420 & 1279 & 1301 & 13000 & 1301 \\
\hline
SNIIn & 10323 & 1359 & 1318 & 13000 & 1318 \\
\hline
SNIIb & 9882 & 1233 & 1208 & 12323 & 1208 \\
\hline
TDE & 9078 & 1162 & 1114 & 11354 & 1114 \\
\hline
SLSN-I & 10285 & 1322 & 1273 & 12880 & 1273 \\
\hline
AGN & 8473 & 1046 & 1042 & 10561 & 1042 \\
\hline
CaRT & 0 & 0 & 10353 & 10353 & $11 \pm 3$ \\
\hline
KN & 0 & 0 & 11166 & 11166 & $11 \pm 3$ \\
\hline
PISN & 0 & 0 & 10840 & 10840 & $11 \pm 3$ \\
\hline
ILOT & 0 & 0 & 11128 & 11128 & $10 \pm 3$ \\
\hline
uLens-BSR & 0 & 0 & 11244 & 11244 & $10 \pm 3$ \\
\hline

\multicolumn{6}{l}{$^a$ The mean number of transients across the 50 test samples is shown. The}\\
\multicolumn{6}{l}{errors refer to the STD in the population size across the 50 sets. All common}\\
\multicolumn{6}{l}{ test data is part of every sample, hence errors are not shown.}\\
\end{tabular}

\end{center}
\end{table}

\subsection{Preprocessing}

To preprocess our light curves, we first remove observations that have a S/N ratio (signal-to-noise) less than 1 (where the noise is simulated based on the observing properties of the ZTF) as a rough threshold for observations that are not meaningful. We correct our light curves for Milky Way extinction, which is solely based on the positioning of the observation in the sky and hence is available in real time. We then define the \textit{trigger} as the first measurement in a light curve that exceeds a $5\sigma$ S/N ratio. We remove all measured observations 70 days after the trigger and 30 days before the trigger as these are likely not part of the transient phase of the light curve. Next, we correct all observation times for cosmological time dilation and scale the times between $-30 < T_{\mathrm{trigger}} < 70$ days to values between 0 and 1. The scaled time is computed as follows,
\begin{equation}
    t = \frac{(T_{\mathrm{obs}} - T_{\mathrm{trigger}}) + 30}{100(1+z)}
\end{equation}
where $z$ is the spectroscopic host redshift and $T$ refers to the observer frame time in Modified Julian Days (MJD). We found that directly incorporating time dilation through the spectroscopic redshift improved our results. However, we acknowledge that the host redshift may not always be available and discuss this limitation briefly in \S \ref{sec:ClassifierMethod}.
We then compute the scaled flux ($f$) and flux error ($\epsilon$) values for each transient by dividing the measured flux and flux error by $500$, a reasonable constant close to the mean flux in our dataset that helps keep the inputs to the neural network small without using any future observations. In this work, we aim to detect anomalies in real-time, and thus we cannot scale the light curves by the peak flux or any method that uses future observations.

We set aside 80\% of the data from the common transient classes for training, 10\% for validation, and 10\% for testing. We use 100\% of our anomalous data for testing and ensure that it is unseen by our model, as mentioned in \S \ref{sec:curse-of-anom}. The number of transients in our dataset from each class are listed in Table \ref{table:populationdistribution}.

\section{Methods}
\label{sec:Methods}
\subsection{Overview and Rationale}

\begin{figure*}
    \centering
    \includegraphics[width=0.8\textwidth]{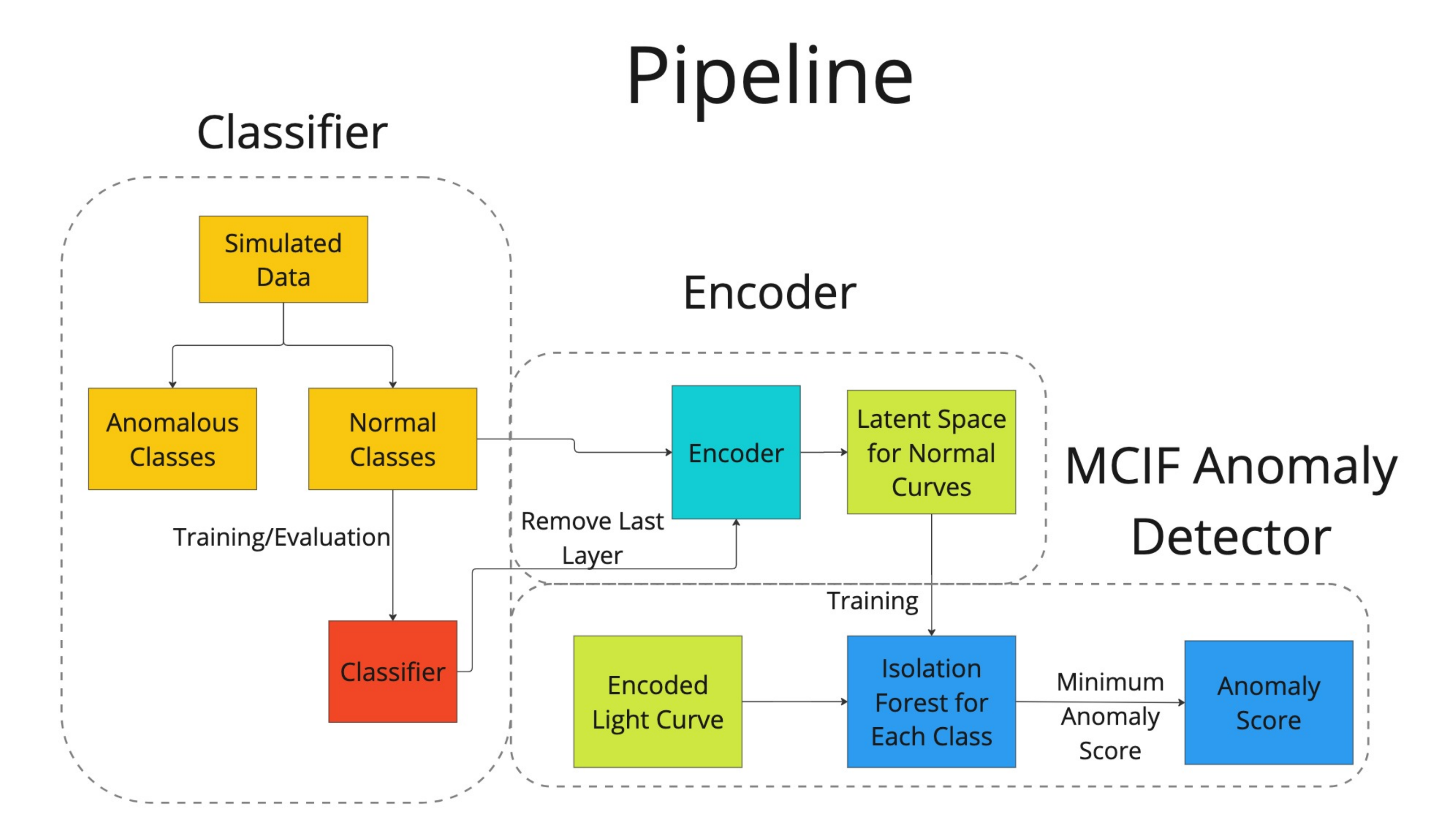}
    \caption{A visual summary of the architecture described in this work. Our approach first trains a classifier, then repurposes it as an encoder, and finally applies Multi-Class Isolation Forests (\texttt{MCIF}), proposed in this work, for anomaly detection. Colors in the plot have changed.}
    \label{fig:Flowchart}
\end{figure*}

Figure \ref{fig:Flowchart} summarizes our methodology. First, we train a Recurrent Neural Network (RNN) to classify the common classes of transients. Then, we remove the final layer of the trained model and use the remaining architecture as an encoder. The penultimate layer of our classifier has 100 neurons, but we find that any reasonably large latent space size works well for anomaly detection (See Section \ref{sec:hyperparameter}). To effectively extract anomalies from a well-represented space, it is essential to ensure that transients from similar classes cluster together. In our encoder, the latent space is directly used for light curve classifications, which should naturally lead to clustering of similar transients.

Once we have established this representation space, we must extract anomalies from it. However, when dealing with multiple clusters, a single isolation forest may struggle to capture each cluster equally (for further details, refer to Appendix \ref{sec:MCIF_Advantages}). This challenge motivated our approach, \texttt{MCIF}, where we train an isolation forest for each class, representing a distinct cluster, and select the minimum anomaly score as the final score. This minimum score should come from the cluster to which the latent observation is closest, providing the desired functionality.

In this work we chose to use a neural-network based architecture for anomaly detection. One of the advantages of using a neural network-based architecture over hand-selected features is that it is a data-driven model, which should make it more sensitive to identifying out-of-distribution data. This inherent quality of neural networks makes them especially good for anomaly detection.

\subsection{Classifier}

\label{sec:ClassifierMethod}
We train a DNN (Deep Neural Network) classifier that maps a matrix of multi-passband light-curve data $\bm{X}_s$ for a transient $s$ to a $1 \times N_c$ vector of probabilities, reflecting the likelihood of the given light curve being each of the aforementioned non-anomalous transient classes, where $N_c$ is the number of classes.

The transient classifier utilizes a Recurrent Neural Network (RNN) with Gated Recurrent Units \citep[GRU,][]{GRU} to handle the sequential time series data. The input for each transient, $\bm{X}_s$, is a $4 \times N_T$ matrix where $N_T$ is the maximum number of timesteps for any input sample. $N_T$ is $656$ in this work, but most transients have much fewer observations. Each row of the input matrix is composed of the following vector,
\begin{equation}
     \bm{X}_{sj} = [f_{sj}, \epsilon_{sj}, t_{sj}, \lambda_p],
\end{equation}
where $f_{sj}$ is the scaled flux for the $j$th observation of transient $s$, $\epsilon_{sj}$ is the corresponding scaled uncertainty, $t_{sj}$ is the scaled time of when the measurement was taken, and $\lambda_p$ is the central wavelength of the passband from which the measurement comes from. For the two passbands in ZTF, the central wavelengths used are as follows \{$\lambda_g=0.4827 \mathrm{\mu m}$, $\lambda_r=0.6233 \mathrm{\mu m}$\}. We include flux error as an input to the DNN to enable it to learn how to weigh individual flux measurements for more accurate classifications. To implement the variable length input in \texttt{tensorflow}, we use a Masking layer and pad the arrays with zero entries to make the input matrix as long as the longest light curve (which is of size $N_T$).

\begin{figure*}
\centering
\includegraphics[width=0.7\textwidth]{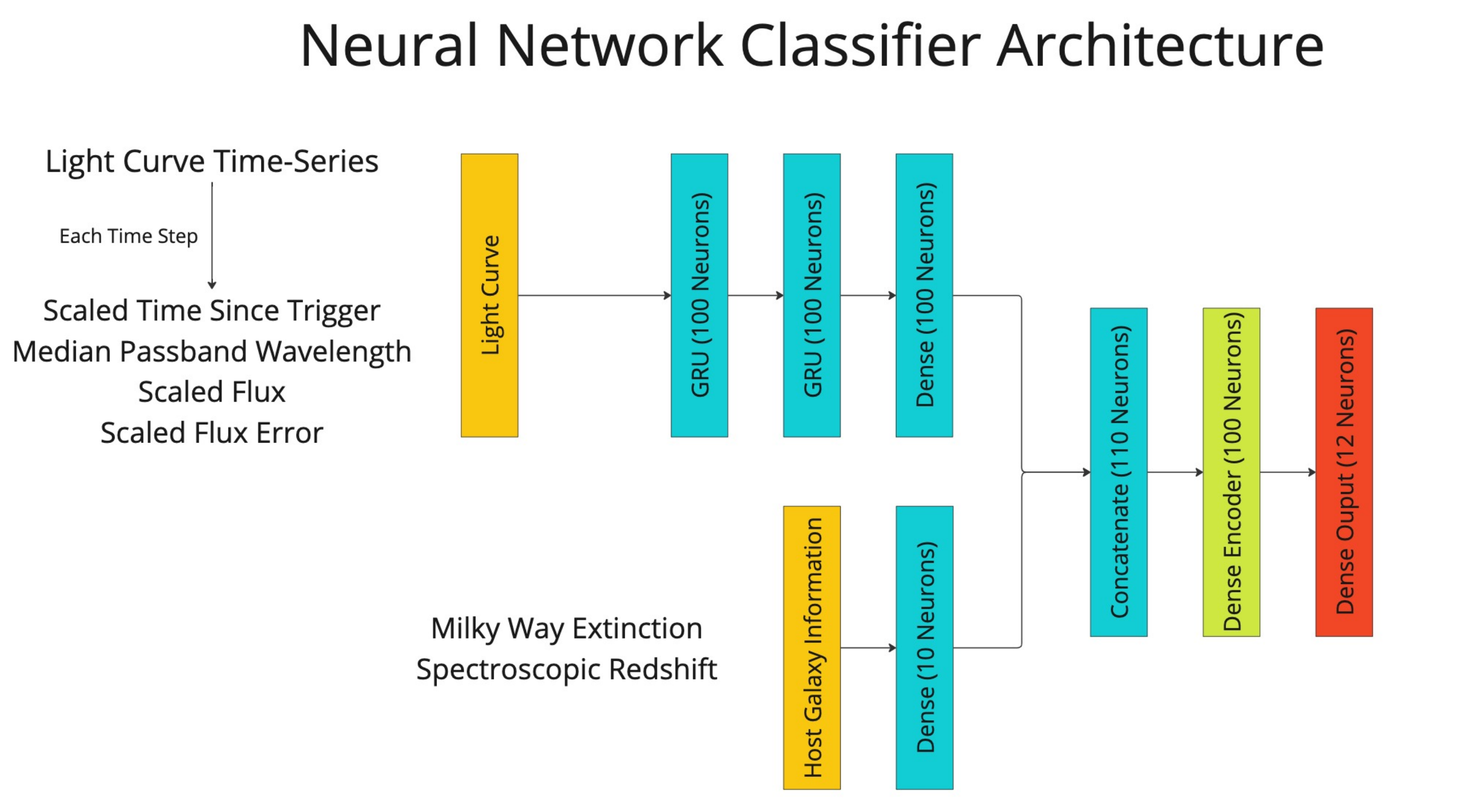}
\caption{A visualization of the neural network classifier being used in this work. Our model has two input streams, one for real-time light curve data and the other for contextual information. The light curve data (first input stream) goes through multiple GRU layers and then a dense layer. The contextual information (second input stream) feeds through a dense layer. The final dense layers from both input streams are merged into a concatenate layer. We feed that to a 100-neuron dense layer that will serve as the latent space of the encoder. Finally, this dense layer feeds into the output layer which provides classification scores. Colors in the plot have changed}
\label{fig:ClassifierOutline}
\end{figure*}

This input format has two major advantages. Firstly, it eliminates the need for interpolation methods to pass sequential data into our model. Interpolation or imputation is often needed in astronomical transient classifiers as observations are recorded at irregular intervals. For real-time light curve tasks, linear interpolation is sometimes used \citep[e.g.][]{OriginalPaper} but yields poor results for sparse light curves or surveys with larger cadences, such as LSST. Linear interpolation may confuse models, particularly when applied to a light curve with prolonged intervals of missing data. In such cases, extended sequences of the interpolated light curve may represent only two original observations.

Gaussian Processes (GP) interpolation has shown success for light curve data (e.g. \citealp{avacado} and \citealp{vraenn}) but does not help with real-time anomaly detection as it requires the whole light curve to perform the interpolation. \cite{vraenn} gives justification for using GP for real-time anomaly detection, stating that each new observation heavily anchors the GP and that the GP is similar to physical model priors. However, methods that avoid any reliance on future observations are more appropriate for real-time analysis and better suited for early anomaly detection.

The second advantage of our input method is the inclusion of wavelength information to represent different passbands. Previous works have typically passed the light curves from different passbands separately, resulting in even larger sequences of missing (thus interpolated or imputed) data in the common case where some passbands have significantly fewer observations than other passbands. In our work, the inclusion of the central wavelength helps the model learn the relationship between different passbands, infer parts of the light curve in bands where few observations are present, and allows for all data to be passed in one channel.

From ZTF, we only get data in the $g$ and $r$ passbands which makes the difference in passbands less significant. However, LSST and other large-scale transient surveys will have data in up to six different passbands, and giving the model insight into relationships between different passbands will be crucial. This learned relationship, along with some transfer learning \citep{Iman_2023} , may make it possible to consolidate data from multiple observatories with different passbands to train one model or allow for a model trained on data from one observatory to be quickly used on data from another observatory.

After the recurrent layers of the DNN, we pass some contextual information into the classifier, which has been shown to be helpful for light curve classification \citep{HostGalInfo}. In this work, we use the Milky Way extinction and the host galaxy's spectroscopic redshift as additional inputs to the network. However, we understand that spectroscopic redshift is not always available, and future work should include training a model with photometric redshift or without redshift entirely.


Figure \ref{fig:ClassifierOutline} illustrates the architecture of the neural network classifier. The classifier was implemented using \texttt{keras} and \texttt{tensorflow} \citep{keras,tensorflow}. We detail the activation functions used in each layer of our classifier as follows. The input stream that each layer is part of is shown in parentheses.

\begin{enumerate}
    \item \textit{Input Layer 1} (Light Curve Stream) - Takes a matrix of shape $4$ x $N_T$ as input to the Recurrent Neural Network.
    \item \textit{Gated Recurrent Unit} \citep{GRU} (Light Curve Stream) - Two recurrent layers consisting of 100 gated recurrent units with tanh activation functions.

    \item \textit{Dense} (Light Curve Stream) - A dense layer consisting of 100 neurons with tanh activation functions. 

    \item \textit{Input Layer 2} (Contextual Stream) - Takes a vector of length $2$ containing the Milky Way extinction and spectroscopic redshift. 
    
    \item \textit{Dense} (Contextual Stream) - A dense layer consisting of 10 neurons with ReLU activation functions. This dense layer is connected to Input Layer 2.

    \item \textit{Concatenate Layer} - A layer to merge the 2 streams of input. This layer concatenates the final dense layers for each input stream into one layer with 110 neurons.
    
    \item \textit{Dense} - A layer with 100 neurons to act as the latent representation for the light curves with a ReLU activation function.

    \item \textit{Dense} - A layer with 12 neurons (1 for each common transient class). This layer has a softmax activation function to map the output values to a probability score.

\end{enumerate}

In our final model architecture, we use GRU layers (Gated Recurrent Units) proposed and tested in \citet{GRU}. They are shown to perform better than typical Recurrent Neural Networks (RNNs) and have quicker training times than Long Short-Term Memory networks (LSTMs; \citealt{LSTM}), another variant of RNNs \footnote{We empirically find that there is little difference between an LSTM and GRU model, in both classification accuracy and anomaly detection.} \citep{GRUvsLSTM}.

To counteract imbalances in the distribution of classes in the dataset, we use a weighted categorical cross-entropy (see equation 6 of \citealt{Muthukrishna19RAPID}) as a loss function with the weight $w_{c}$ proportional to the fraction of transients from each class $c$ in the training set,

\begin{equation}
    w_{c} = \frac{T}{T_{c}}, 
    \label{eq:1}
\end{equation}

where $T_c$ is the number of transients from the class $c$ and $T$ is the total number of samples in the training set. This weighting scheme ensures that classes with fewer samples have higher weights. To train the classifier, we ran it over $40$ epochs using the Adam optimizer \citep{adam} with \texttt{EarlyStopping} implemented in \texttt{keras}.

\subsection{Multi-Class Isolation Forests}

\label{sec:AnomalyMethod}

Once the classifier is trained, we remove the last layer and use the remaining architecture to map any light curve to the latent space. We define this encoder as a function $E(\bm{X}_s)$, that takes the aforementioned preprocessed light curve data, $\bm{X}_s$, and maps it to a 100-dimensional latent space $\bm{z}_s$,

\begin{equation}
    \bm{z}_s = E(\bm{X}_s).
\end{equation}

For anomaly detection, we now want to compute the anomaly score,
\begin{equation}
    a_{s} = A(\bm{z}_s),
\end{equation}

where $A(\bm{z}_s)$ is a function that evaluates the anomaly score $a_s$ for a latent observation $\bm{z}_s$. The goal of this work is to generate relatively large anomaly scores for anomalous transients and smaller anomaly scores for non-anomalous transients. 

Isolation forests are known to be a very simple yet effective anomaly detection algorithm, especially in the domain of astronomical time series \citep[e.g][]{vraenn, Ishida2021_Timeseries, IsoForestUse2}. However, using a single isolation forest performs poorly in determining some common classes as non-anomalous. This challenge arises from the complexity of our latent space, which contains various distinct clusters that pose difficulties for a single isolation forest to differentiate. Thus, we propose a new framework, \texttt{MCIF}, where an isolation forest is trained separately on data from every class, using the minimum anomaly score from any isolation forest as the final anomaly score.

We define $12$ isolation forests, $I_c(\bm{z}_s)$, trained on latent space observations from the common transient class $c$. The final anomaly score is defined as

\begin{equation}
    A(\bm{z}_s) = \min_{\forall c} \Bigl( -I_c(\bm{z}_s) \Bigr).
\end{equation}

The function $I_c(\bm{z}_s)$ is positive for less anomalous transients and negative for anomalous ones, to be consistent with the \texttt{sklearn} implementation of isolation forests. We negate the scores as we prefer defining transients with higher anomaly scores to be more anomalous, but this makes no difference to the results. All isolation forests used in this work are trained with 200 estimators. The results of using a single isolation forest and the benefits of using Multi-Class Isolation Forest are explored further in Appendix \ref{sec:MCIF_Advantages}.

\subsection{Limitations of Evaluating Anomaly Detection Methods}

\label{sec:curse-of-anom}

Evaluating the performance of anomaly detection models is challenging because anomalies are rare, and it is difficult to build validation sets that account for the unknown. To simulate seeing \textit{anomalous} data for the first time, the five aforementioned anomalous classes are not revealed to the model until final evaluation. This approach mimics the real-world situation where anomaly detectors (and astronomers) have limited knowledge of the anomalies they may discover. 

The goal of this work is not to identify the specific anomalous classes mentioned, but rather to detect anomalies in general. Hence, using physical model priors or giving the model any information about anomalous classes beforehand would reveal too much about the specific rare transient classes used in this work. This could hinder the model's performance on novel, real-world anomalies that may differ from the ones used in this study.

\section{Results and Analysis}
\label{sec:Results}

In this section, we first evaluate the performance of our neural network classifier on distinguishing between common transient classes (\S \ref{sec:ClassifierRes}). We then analyze how well our proposed anomaly detection pipeline, utilizing the classifier as an encoder (\S \ref{sec:LatentRepresentation}), is able to identify rare/anomalous transients (\S \ref{sec:AnomalyRes}).

\subsection{Classifier}
\label{sec:ClassifierRes}

\begin{figure*}
\centering
\begin{tabular}{ll}
\includegraphics[width=0.5\textwidth]{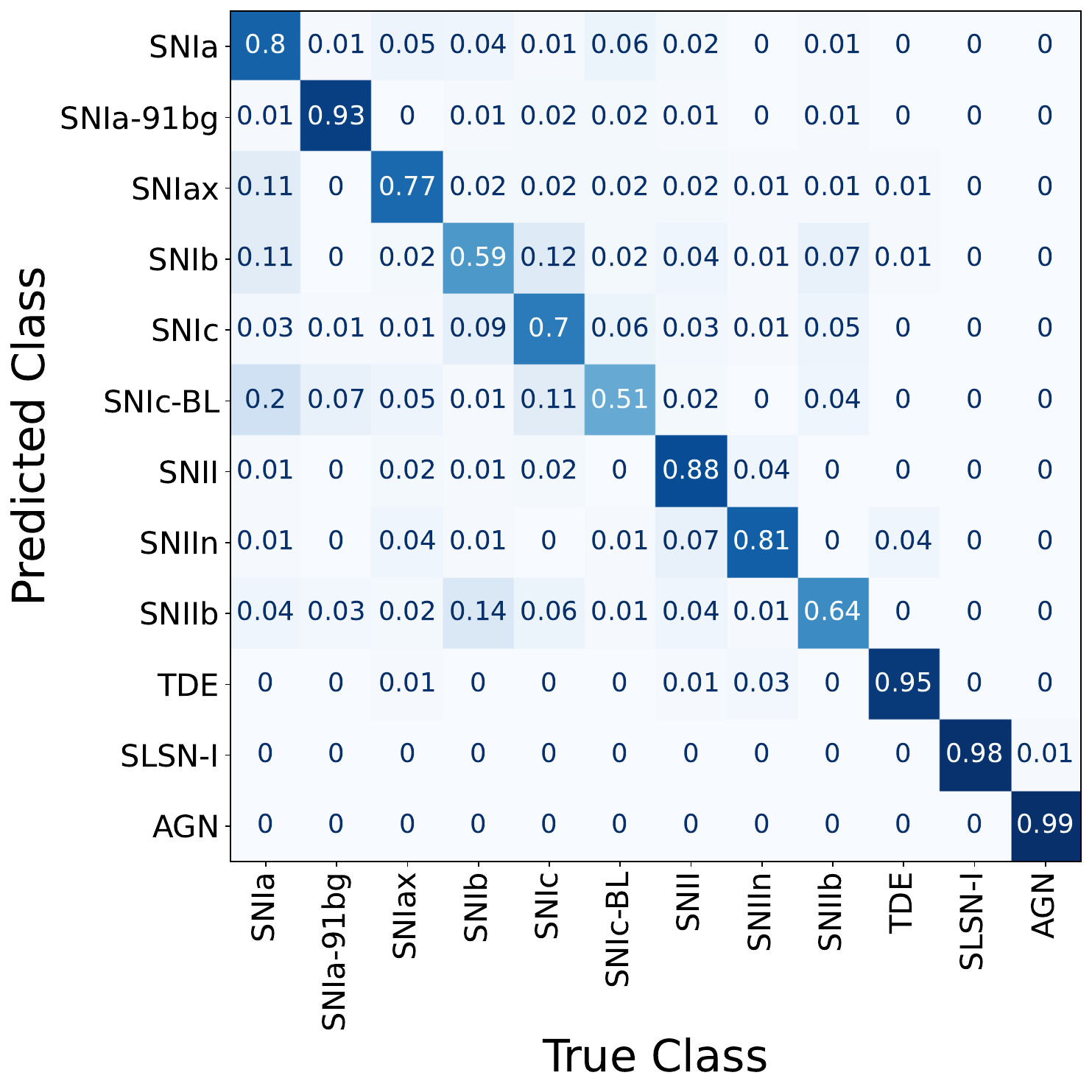}
&
\includegraphics[width=0.5\textwidth]{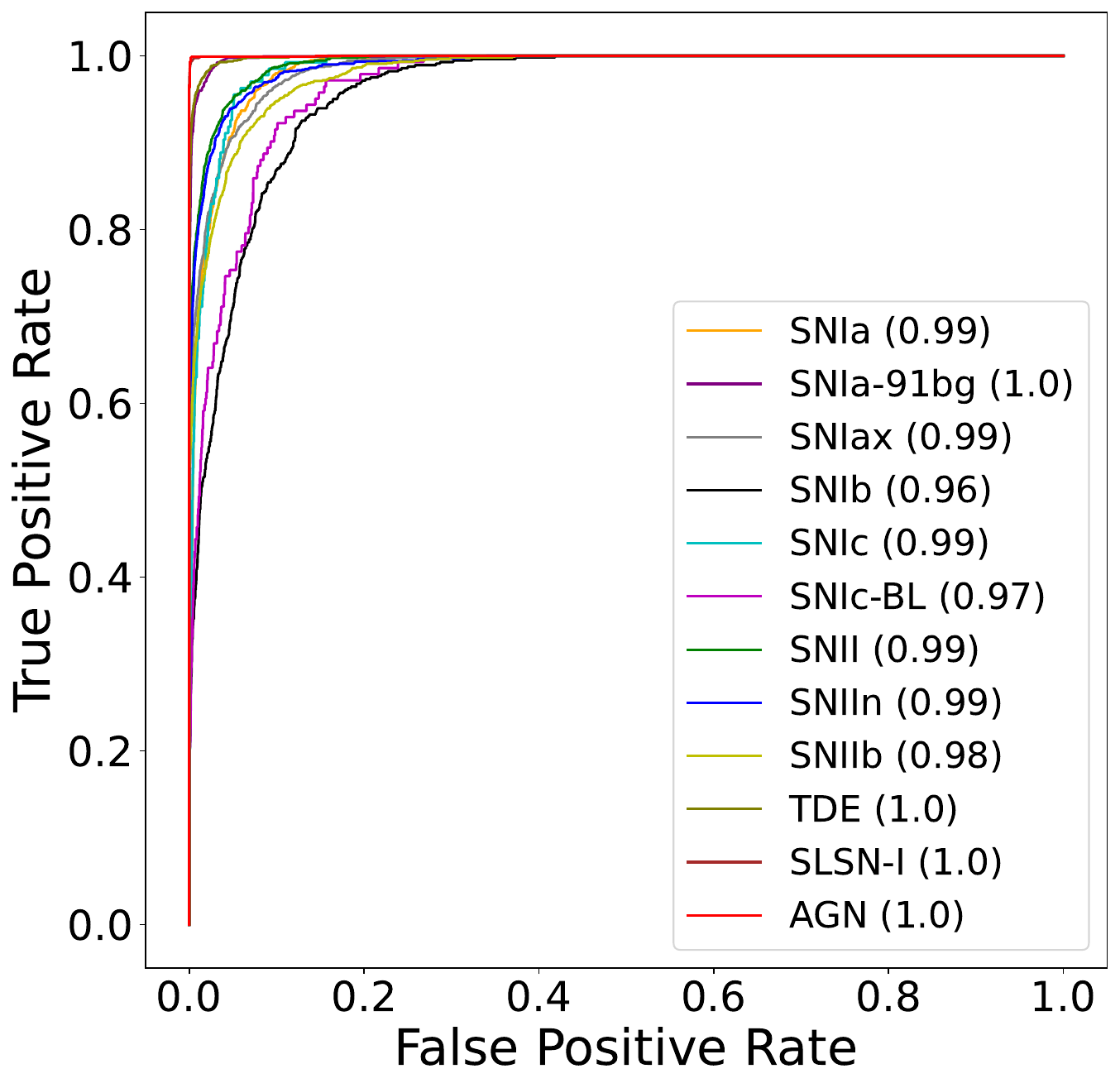}
\end{tabular}
\caption{The normalized confusion matrix [left] and ROC curve [right] of the 12 common transient classes used for training given full light curve data. Each cell in the confusion matrix signifies the fraction of transients from each \textit{True Class} that was classified into the \textit{Predicted Class}. The ROC curve illustrates the True Positive Rate against the False Positive Rate across various threshold probabilities for each class, with the Area Under ROC curve (AUROC) in parenthesis. The model's evaluation is conducted on the test set consisting of 10\% of the data from the common classes.}
\label{fig:ConfusionROC}
\end{figure*}

In this section, we evaluate the performance of our classifier on the test set consisting of the 12 common transient classes.

The normalized confusion matrix in Figure \ref{fig:ConfusionROC} [left] illustrates the classifier's ability to accurately predict the correct transient class on the test data. Each cell indicates the fraction of transients from the true class that are classified into the predicted class. The high values along the diagonal, approaching 1.0, indicate strong performance.  The misclassifications, indicated by the off-diagonal values, predominantly occur between subclasses of Type Ia supernovae (SNIa, SNIa-91bg and SNIax) and between the core-collapse supernova types (SNIb, SNIc, SNII subtypes), which is expected given their observational similarities. These SNe have been shown to confuse previous models (see Fig. 7 of \citealp{Muthukrishna19RAPID}).

While the confusion matrix provides valuable insights into the accuracy of our model, it only considers the highest-scoring predicted class and does not use the continuous probability scores that our classifier outputs for each possible class. The Receiver Operating Characteristic (ROC) curve, shown in Figure \ref{fig:ConfusionROC} [right], effectively uses these probabilities. It plots the True Positive Rate (the fraction of positive samples correctly identified as positive) against the False Positive Rate (the fraction of negative samples incorrectly identified as positive) for each class across a range of threshold probability values. This metric is particularly useful in a multi-class context, as it captures the model's ability to assign low probabilities to several classes when it is uncertain. A key measure of performance, the Area Under the ROC Curve (AUROC), quantifies the overall ability of the classifier to discriminate between classes. In our study, high AUROC values, approaching 1 for all classes, underscore the robustness of our classifier.

A brief comparison of the results in this work and those of a similar DNN light-curve classifier \citep[\texttt{RAPID};][]{Muthukrishna19RAPID} hints that our classifier can perform on par with a reasonable baseline and has better accuracies for many classes. This improvement can be attributed, in part, to the use of an improved simulated dataset that has fixed some problems with the simulation of several events including a lack of diversity of core-collapse supernovae (see \citealt{Vincenzi2021}, for details on the problems with the original PLAsTiCC simulations). \texttt{RAPID} also included some of the rare classes that we deliberately did not include in our classifier and instead designated as anomalous (See Fig. 7 of \citealp{Muthukrishna19RAPID}). While our classification methodology was very similar to \texttt{RAPID}, a key difference was our unique input method that bypassed the missing data problem. Future work should compare the effectiveness of this input method with traditional interpolation or imputation methods, and further comparison is beyond the scope of this work.

\subsection{Latent Representation}
\label{sec:LatentRepresentation}

\begin{figure*}
\centering
\begin{tabular}{ll}
\includegraphics[width=0.5\textwidth]{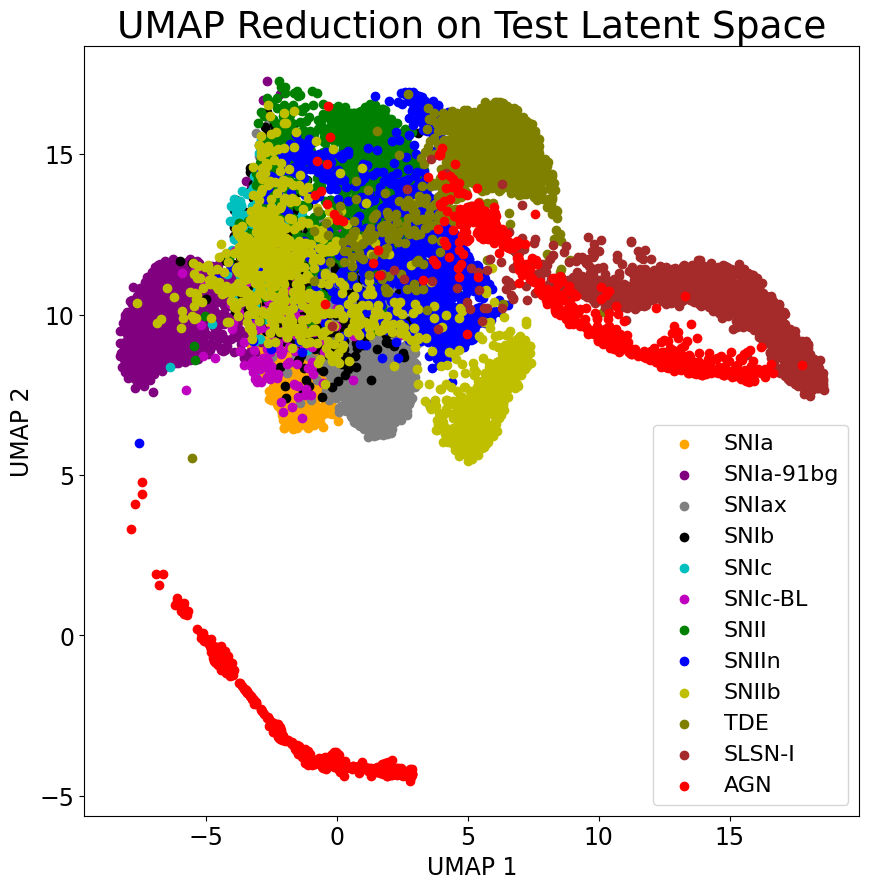}
&
\includegraphics[width=0.5\textwidth]{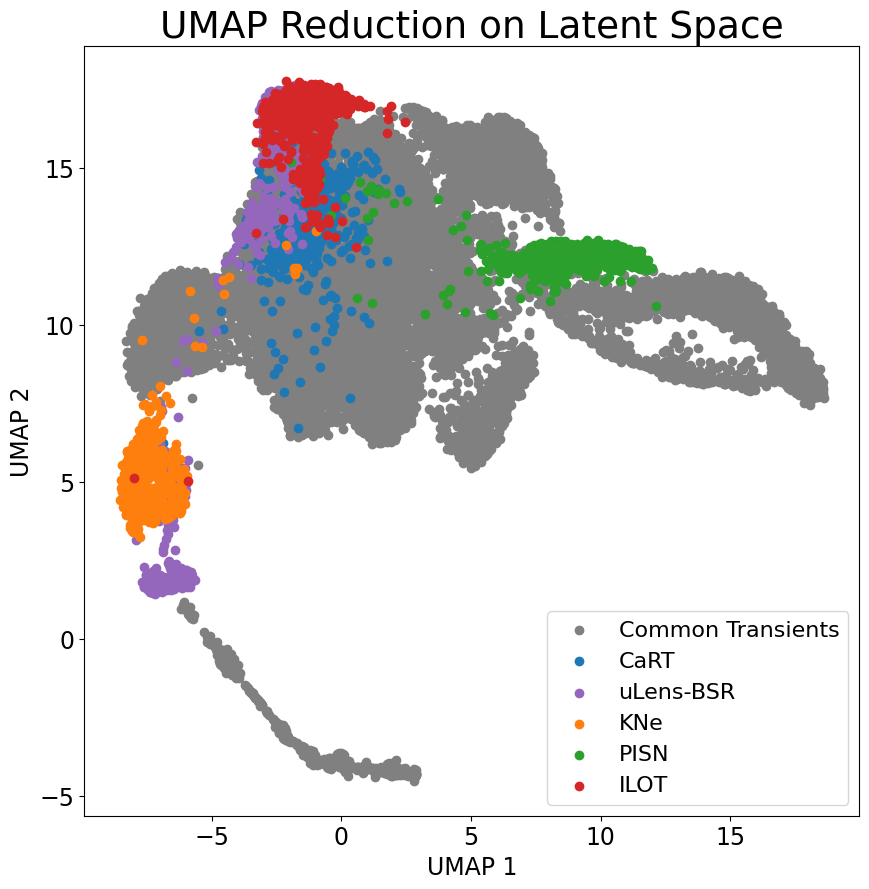}
\end{tabular}
\caption{The UMAP reduction of the latent space derived from the test set, which includes 10\% of the common transients reserved for testing the classifier [left] and randomly sampled anomalous transients from the unseen anomaly dataset [right]. Despite not being trained on this data, the learned features still exhibit clear visual structure and anomalous transients form distinct clusters separate from the common classes. It is important to note that the UMAP reduction is used only for visualization purposes, and the actual anomaly detection (as seen in Figure \ref{fig:MCIFAverageScore} and the remaining plots) is performed on the 100-dimensional latent space.}
\label{fig:UMAP}
\end{figure*}

After repurposing the classifier as an encoder, we obtain a 100-dimensional latent space. We can visualize this latent space with UMAP \citep{UMAP}, a manifold embedding technique, to determine if there is visible clustering\footnote{We use the \texttt{umap-learn} implementation in \texttt{python} using the hyperparameters ``minimum distance'' set to 0.5 and "number of neighbors" set to 500.}. In Figure \ref{fig:UMAP} [left], we plot the UMAP representations of the test data. While it is difficult to examine some of the overlapping classes in this embedded space, there is clear clustering of many of the classes.
In Figure \ref{fig:UMAP} [right], we color all of the common classes grey and include a sample of transients from the anomalous classes. We see that the anomalous classes cluster together in the embedded space and separate from the common transients despite the model not being trained on these objects. This level of clustering suggests that our encoder may be discovering generalisable patterns within light curves, and this property may have potential use cases beyond anomaly detection in few-shot classification.
It is important to note that we only use UMAP for visualisation purposes and that the latent space used for anomaly detection is obtained directly from the penultimate layer of the classifier.

\begin{figure*}
\centering
\includegraphics[width=0.8\textwidth]{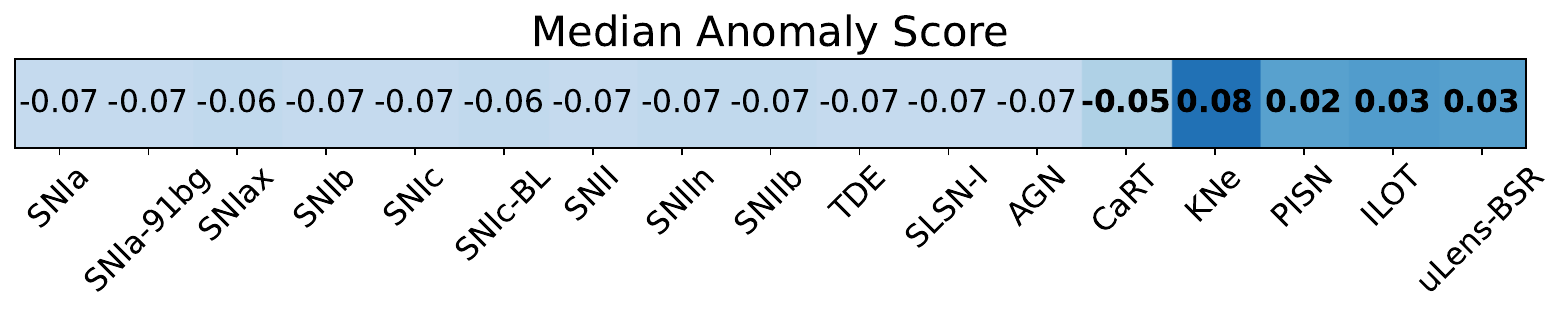}
\caption{The median anomaly score (rounded to two decimal places) for each class extracted from the latent representations of full light curves. The scores come from the full, unseen anomalous dataset for anomalous classes and the 10\% test dataset for common classes. The five classes on the right (in bold) are anomalous. The separation between the scores of anomalous classes and common classes is evident, and the anomaly scores for the common classes are consistently low signifying they are not erroneously marked anomalous. For further analysis, Figure \ref{fig:Distribution} shows the full anomaly score distribution for each class.}
\label{fig:MCIFAverageScore}
\end{figure*}

\subsection{Anomaly Detection}
\label{sec:AnomalyRes}
After training \texttt{MCIF} on the latent representations of the training data, we pass unseen test data and anomalous data through our pipeline for evaluation. In Figure \ref{fig:MCIFAverageScore}, we list the median anomaly score predicted by \texttt{MCIF} for each class. The anomalous classes have much higher median anomaly scores than the common classes, illustrating a significant distinction between the scores of all common and most anomalous classes. This difference is not as pronounced when using a single isolation forest and the advantages of employing \texttt{MCIF} are discussed further in Appendix \ref{sec:MCIF_Advantages}.

\begin{figure}
\centering
\includegraphics[width=0.475\textwidth]{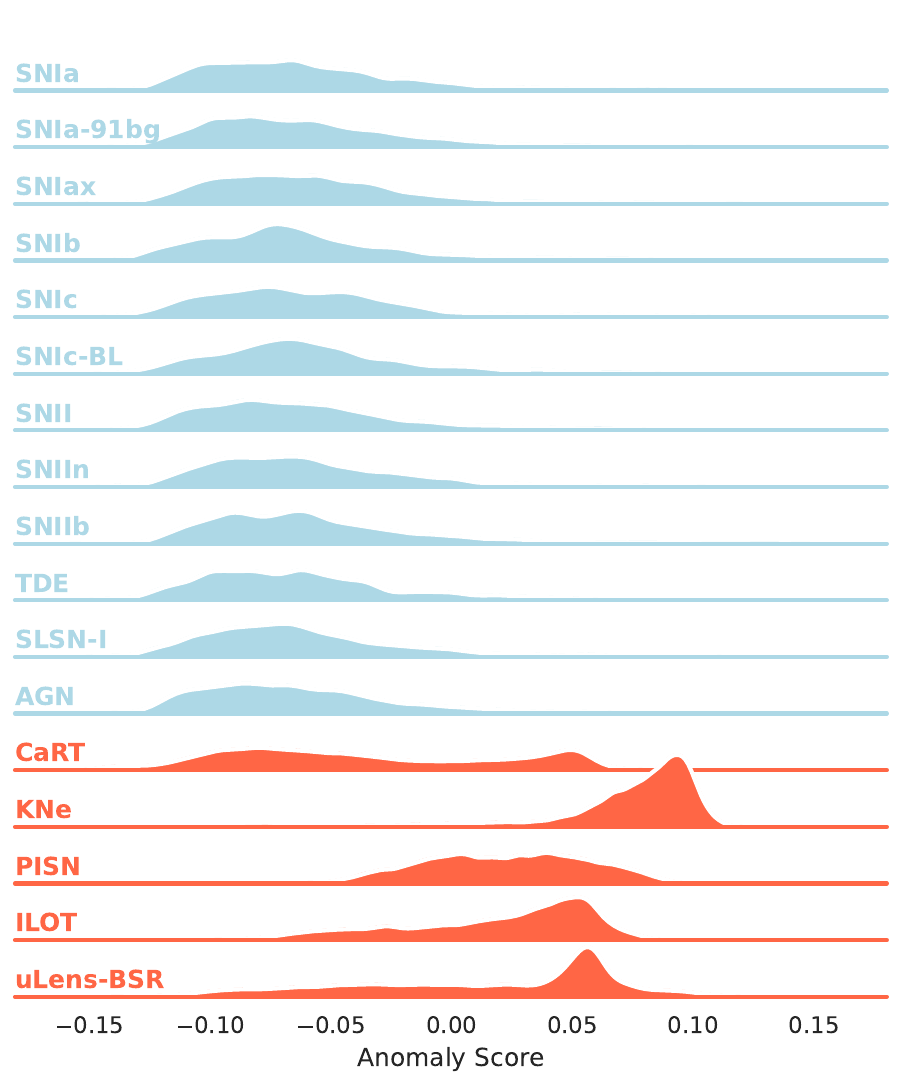}
\caption{The distribution of anomaly scores for each class, computed using \texttt{MCIF} on the latent representations derived from full light curves. The scores are plotted using $100\%$ of the anomalous dataset (unseen during training) and the test dataset of common classes. The anomalous classes (bottom five in red) generally show higher anomaly scores with positively skewed distributions. The common classes and CaRTs all have low anomaly scores on average.}
\label{fig:Distribution}
\end{figure}

In Figure \ref{fig:Distribution}, we plot the distribution of anomaly scores predicted by \texttt{MCIF} for each class. The plot further demonstrates the distinction in anomaly scores of common and anomalous transients. Notably, there is a significant skew towards larger anomaly scores for the anomalous classes, reaffirming our model's performance. However, Calcium Rich Transients (CaRTs), despite being one of our anomalous classes, tend to have lower anomaly scores. CaRTs are notoriously difficult to photometrically classify as anomalous due to their resemblance to other common supernova classes (see Fig. 8 of \citealt{Muthukrishna19RAPID} for example). One of the most effective ways to detect CaRTs is to observe a calcium line in their spectra, and a robust anomaly detector would use photometric differences to discern this spectroscopic dissimilarity. However, ZTF is limited to only the $g$ and $r$ passbands, which lets this subtle spectroscopic difference go unnoticed. The upcoming Legacy Survey of Space and Time on the Rubin Observatory will observe data in six passbands and will likely mitigate this issue.

\subsubsection{Anomaly Precision and Recall}
To identify anomalies with \texttt{MCIF}, a threshold anomaly score would need to be chosen such that the common transient classes are not flagged as anomalous, while only the anomalous classes are flagged as anomalous. This threshold score needs to lead to a high \textit{precision} and a high \textit{recall} of anomalies. Precision is a measure of how pure our anomaly predictions are, and recall is a measure of how many anomalies we can expect to find. 
We define anomaly precision and recall as

\begin{equation}
    \label{eq:prec}
    P_{c, \tau} = \frac{|\{ L^s \in L^c \mid A(L^s) > \tau \}|}{|\{ L^s \in L \mid A(L^s) > \tau \}|}
\end{equation}

\begin{equation}
 \label{eq:rec}
    R_{c, \tau} = \frac{|\{ L^s \in L^c \mid A(L^s) > \tau \}|}{|L^c|}
\end{equation}
where $P_{c, \tau}$ and $R_{c, \tau}$ are the precision and recall of class $c$, the tested class, at threshold anomaly score $\tau$, $L$ is the set of all transients, $L^c$ is the set of all transients from class $c$ in $L$, and $L^s$ is a transient from either $L^c$ or $L$. We further define the set $L$ to be comprised of half class $c$ and half of transients coming from the opposite \textit{type} as $c$ when computing the precision and recall for class $c$. For example, if the tested class was KN, the set $L$ would contain half KN and half common transients. Note that only precision (and not recall) is influenced by the composition of $L$.

In other words, precision is calculated as the number of predicted anomalies from each class divided by the total number of predicted anomalies, and recall is calculated as the number of predicted anomalies from each class divided by the total number of transients from that class. Both are defined given a threshold anomaly score and over a deliberately defined sample (as described above).

\begin{figure*}
\centering
\includegraphics[width=0.7\textwidth]{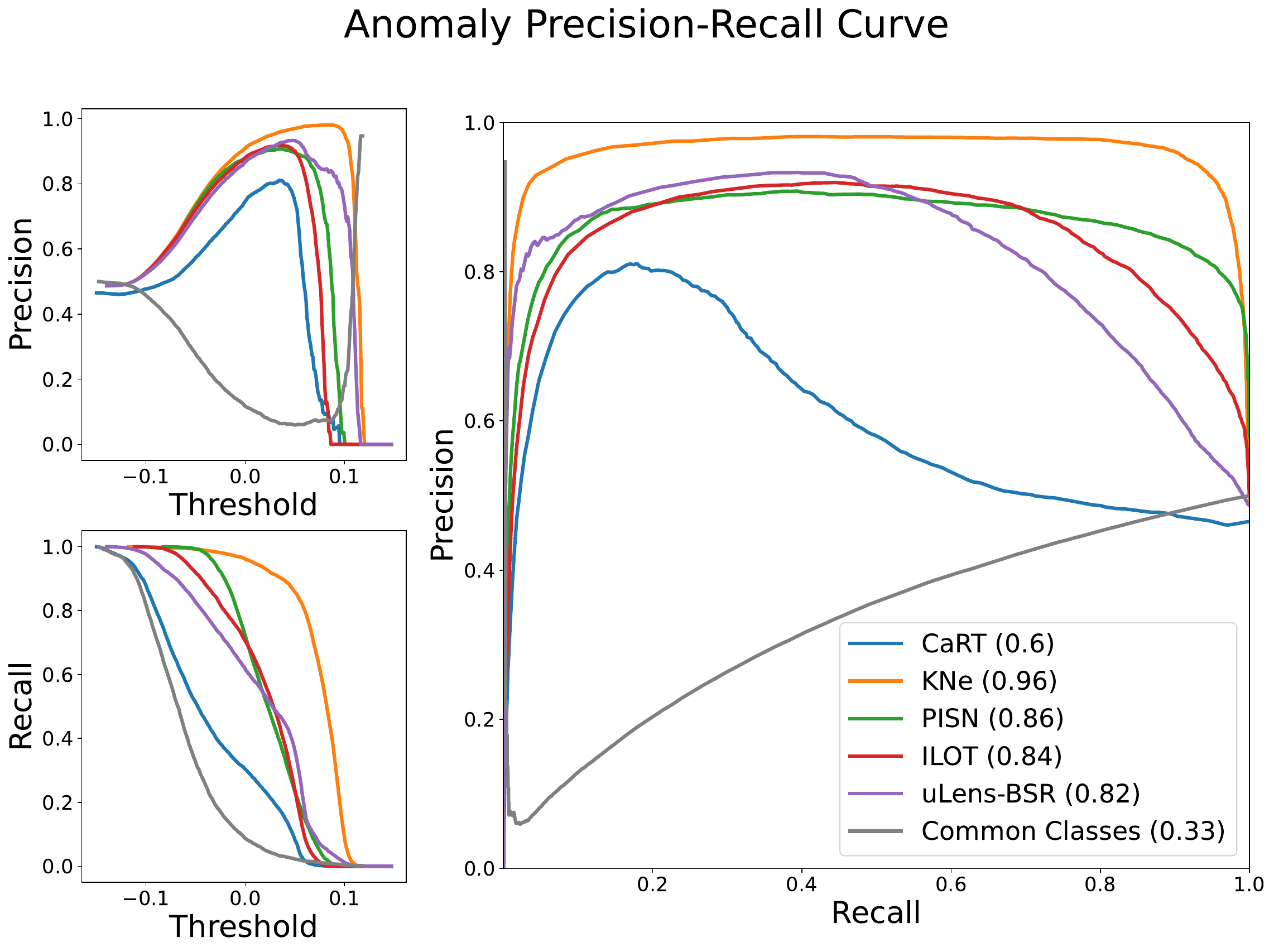}
\caption{The precision-recall curves for our anomaly detection pipeline for each anomalous class and a grey line indicating the average across all common classes. The precision and recall are plotted against the threshold anomaly score in the left sub-figures. Precision and recall are defined in Equation \ref{eq:prec} and Equation \ref{eq:rec}, respectively, and calculated on a set comprising half of the transients from the tested anomalous class and half randomly sampled common transients (all coming from the test data that was unseen by the model). Promisingly, the Area Under the Precision-Recall Curve (AUCPR) for each anomalous class (except CaRTs) is very high. The AUCPR for the common classes is low indicating that they are not often predicted as anomalous by \texttt{MCIF}.}
\label{fig:PrecisionRecallCurveAnom}
\end{figure*}

In Figure \ref{fig:PrecisionRecallCurveAnom} we plot the anomaly precision and recall at various threshold anomaly scores $\tau$ for all anomalous classes and an average for all common classes. In this context, the precision is $0.5$ at the lowest threshold, as at that point all transients are marked as anomalous, and 50\% of them are from the tested class. Recall is $1$ at the lowest threshold, much like for many other machine learning tasks, as all transients of the tested class are identified as anomalous. This interpretation serves as the rationale for the 50-50 composition of the set $L$, as otherwise, it would be difficult to standardize AUC scores across anomalous and common classes.

\begin{figure*}
  \centering
  \begin{tabular}{ll}
  \includegraphics[scale=0.5]{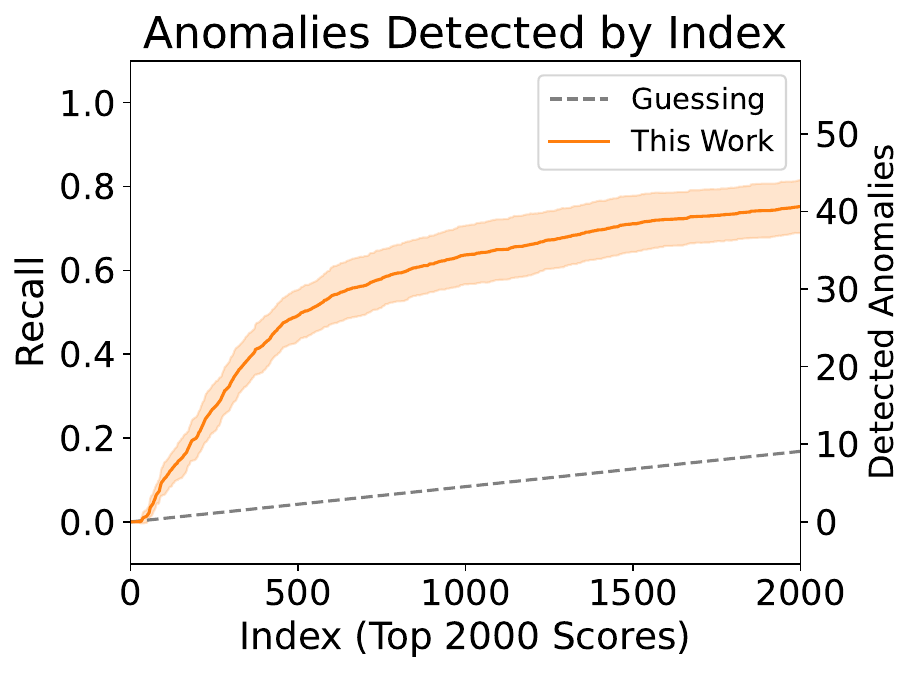}  
  &
  \includegraphics[scale=0.5]{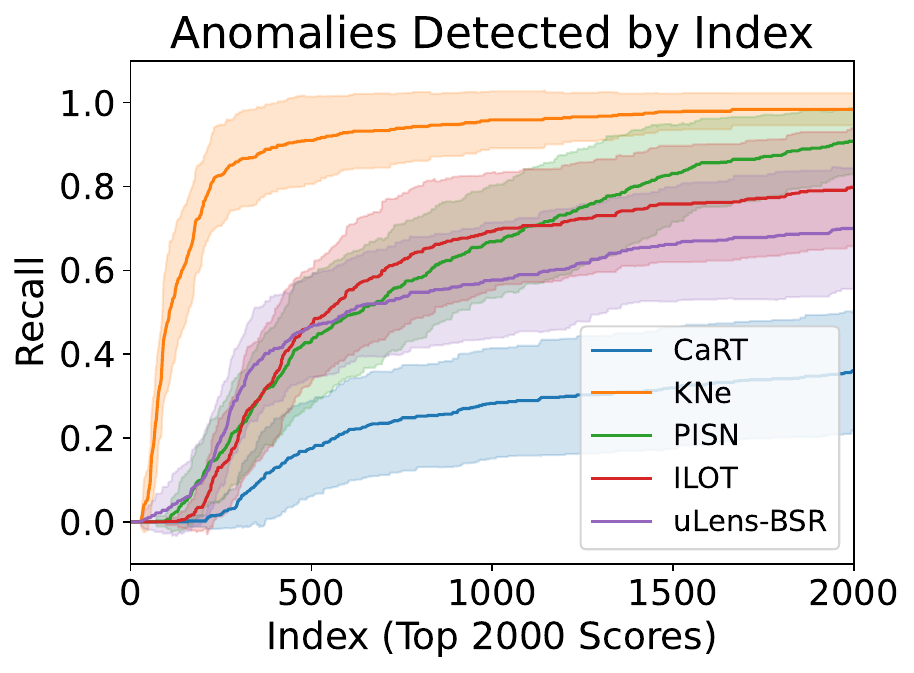}  
  \end{tabular}

  \caption{Anomalies detected in the 2,000 top-ranked transients by \texttt{MCIF} anomaly score index, using a test sample reflecting the estimated frequency of anomalies in nature. In the sample of 12,040 common transients and 54 anomalous transients, the model recalls $41\pm3$ $(\sim75\%)$ of the anomalies after following up the top 2000 ranked transients. The left plot aggregates all anomalies and the right plot delineates per class.
  To control for the variance imposed by the small anomaly sample size, we repeat the sampling 50 times. The mean and standard deviation of detected anomalies are plotted as the solid lines and shaded regions, respectively. }
  \label{fig:Index}
\end{figure*}

The Area Under the Precision-Recall Curve (AUCPR) is a good indicator of how often a class is being marked anomalous. The AUCPR scores for anomalous classes are significantly better than the AUCPR scores for the common classes, even CaRTs. However, the precision of the common classes jumps suddenly at high thresholds while it declines for anomalous classes. This signifies that the transients with the highest anomaly scores are false positive anomalies. This phenomenon is understandable if we consider that most anomalies populate anomaly scores $a_{s} \lesssim 0.1$ in Figure \ref{fig:Distribution}. Beyond this, at thresholds $\tau \gtrsim 0.1$, the recall drops as these anomalous transients no longer meet the threshold. With inherently more common transients, a few extreme latents among them dictate precision initially. Figure \ref{fig:Index} confirms that the top 5-10 candidates are common transients, but after this, as $\tau$ reduces, our model captures a much higher fraction of true anomalies.  

Selecting a threshold anomaly score near the upper-right region of the precision-recall curve will be a good choice for identifying as many anomalies as possible while still having a pure sample with few false positives. This point represents a trade-off between high precision and high recall, often occurring where the curve begins to plateau or shows a notable change in slope.

\subsubsection{Detection Rates in a Representative Population}
\label{sec:representative_population}
The previous results do not take into account that anomalous transients are inherently less frequent than common transients. While the frequency of anomalies in nature is not known, a good estimate for the expected population frequency was presented in \citet{PlasticcSim} for the PLAsTiCC dataset \citep{PlasticcData}. The rate of common transients, as defined in this work, was roughly 220 times larger than that of anomalous transients, using PLAsTiCC frequencies for each class. We used this rate to randomly select a more realistic test dataset that contained 12,040 normal transients and 54 anomalies. Randomly selecting a representative sample of only 54 anomalies is subject to significant variance. Therefore, we created 50 sample datasets to perform 50-fold cross-validation. The mean and standard deviation of the number of transients from each class present in our 50 test samples are listed in Table \ref{table:populationdistribution}.

For each validation set, we ranked the transients by the anomaly scores predicted by \texttt{MCIF}. We followed up the top 2000 ranked transients (roughly 15\% of the dataset) as the candidate pool. Across 50 repeated trials, we identified $41\pm3$ out of the 54 true anomalies in our dataset (recalling $\sim75\%$ of the anomalies). In Figure \ref{fig:Index}, we plot the fraction of anomalies recalled\footnote{This usage of the word \textit{recall} has a different population distribution than defined in Equation \ref{eq:rec}.} and the total number of anomalies recovered for thresholds up to the top 2000 transients. \texttt{MCIF} recalls most anomalies within candidates having the highest anomaly scores, followed by a tapering as fewer anomalies remain. 

Examining the detection rate for each anomaly class, we see that the model's trouble in identifying CaRTs as anomalous brings down the overall anomaly recall. This plot and the precision-recall curves show a consistent pseudo-hierarchy of which anomalies are easiest to detect. If we exclude CaRTs from our sample of anomalies, our recall of anomalies increases to $47\pm2$ out of the 54 true anomalies in our dataset (recalling $\sim87\%$ of the anomalies).

It is worth emphasizing that the objective of this work is to identify anomalies in a general sense rather than tailoring to specific classes, and therefore, this work does not rely on specific information about the anomalous classes defined (see \S \ref{sec:curse-of-anom}). The only specific attribute used is the estimated frequency of anomalies (220 times less frequent than the common classes), which serves as a reference as it is impossible to estimate a similar number for anomalies that have never been observed.

Given the complexity of our deep learning approach, it's important to examine how the sparsity of light curve sampling affects anomaly detection performance. Sparsely sampled light curves from common classes could potentially be assigned large anomaly scores if the model struggles to accurately represent them in the latent space. To investigate this, we analyzed the relationship between the number of observations in a light curve and its likelihood of being classified as anomalous. Our analysis revealed that while anomaly detection performance generally improves with more observations, sparsely sampled light curves are not disproportionately classified as false positives. This resilience to sparse sampling may be attributed to our RNN-based architecture and input method, which are designed to handle irregular time series data. The ability of our model to maintain performance even with limited observations is particularly valuable for early detection of anomalies in ongoing surveys.

\subsubsection{Real-Time Detection}

Identifying anomalies in real-time is important for obtaining early-time follow-up observations, which is crucial for understanding their physical mechanisms and progenitor systems. However, directly assessing our architecture's real-time performance is challenging due to the irregular sampling of light curves in our input format.

\begin{figure*}
  \centering

  \begin{tabular}{ll}
  \includegraphics[scale=0.40]{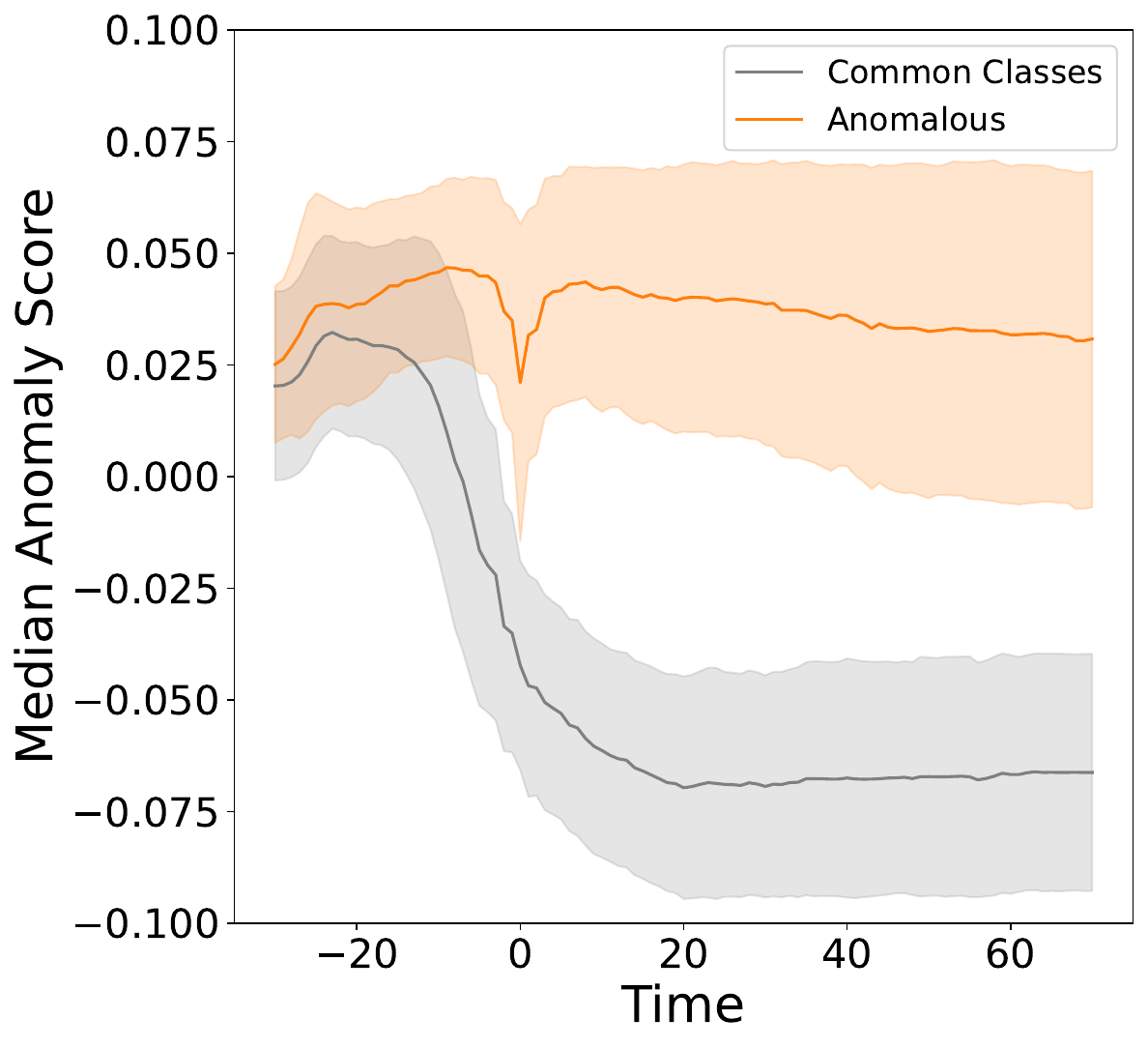}  
  &
  \includegraphics[scale=0.40]{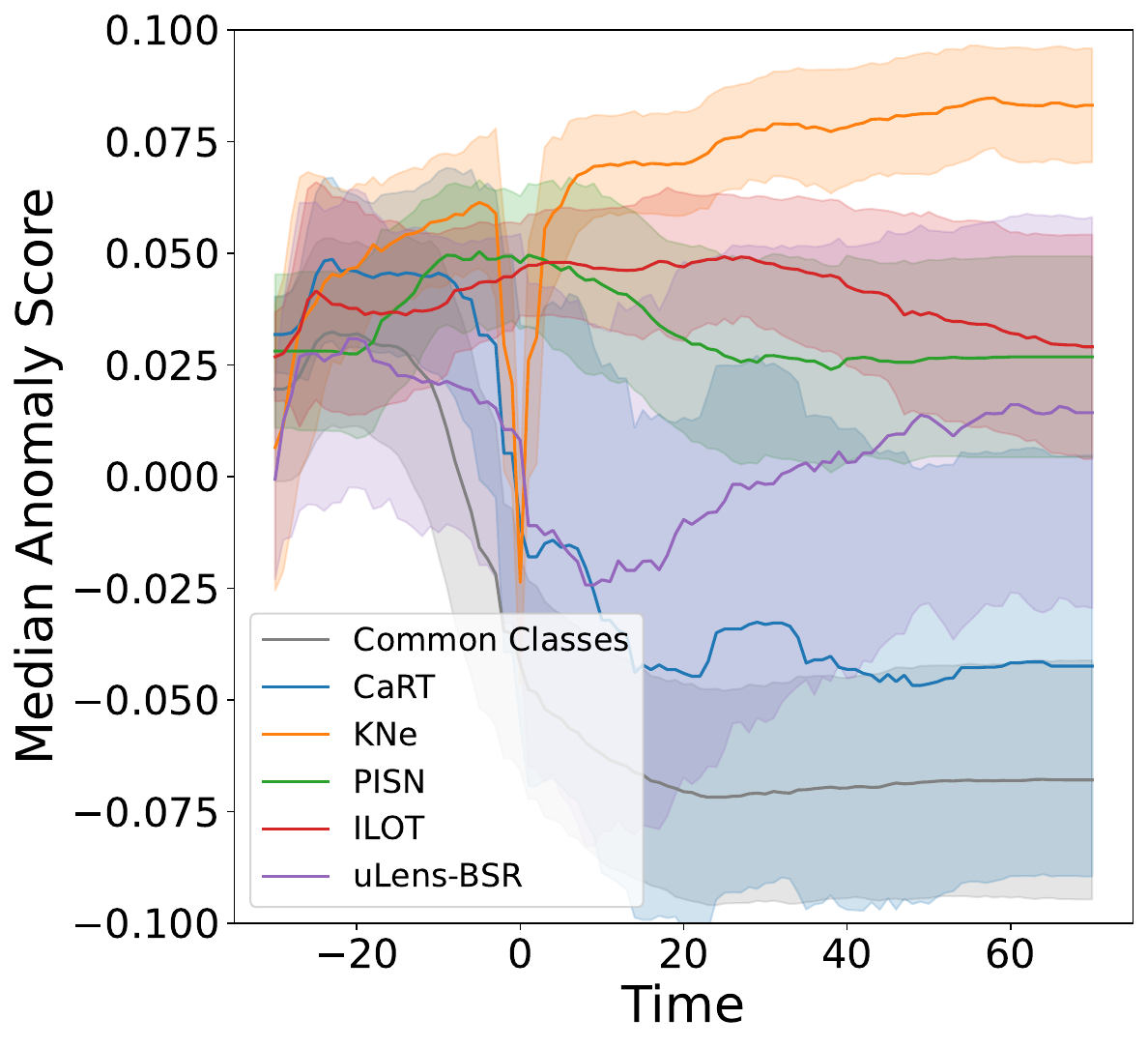}  
  \end{tabular}

  \caption{Median \texttt{MCIF} anomaly score over time for a sample of transients from the test set. Real-time anomaly scores are calculated at intervals of 1 day for a sample of 2000 common and 2000 total anomalous light curves. The left plot shows the scores for the common and anomalous transients as a whole, while the right plot shows each anomalous class individually. The anomaly scores for the common transients decline before the trigger, while the anomalous transients remain at high scores throughout most of the transient's evolution.}
  \label{fig:RealTimeAnalysis}
\end{figure*}

To assess the real-time performance of our architecture, we plot the median anomaly scores over time for a sample of 2000 common and 2000 anomalous transients in Figure \ref{fig:RealTimeAnalysis}. To construct this plot without relying on interpolation, we calculate scores at discrete times $l$ sampled at 1-day intervals from $-30$ to $70$ days relative to trigger, using only observations occurring before each time $l$ to mimic a real-time scenario. To ensure sufficient information for robust scoring, we only consider transients where the final observation was recorded after time $l - 5$. The results show a clear divergence where common transient scores tend to decline around trigger, while anomalous transient scores remain consistently high. 

Figure \ref{fig:RealTimeAnalysis} reveals two notable irregularities. Firstly, the anomaly scores for common transients decline before trigger, which is unexpected given that the pre-trigger phase of most transient classes should primarily consist of background noise. Further analysis of the pre-trigger classification results (Figure \ref{fig:otherealtime}) reveals that certain transients, most notably SLSN-I and AGN, are almost all classified before trigger, thereby lowering the average anomaly score for common transients. This can be attributed to the fact that redshift and pre-trigger information such as host galaxy color and some AGN pre-trigger variability are particularly useful for classifying these transients before trigger (see Figure 16 of \citealp{Muthukrishna19RAPID}).

Secondly, KN exhibit a significant dip around the time of trigger. Upon further analysis, we found that certain common transient classes also experienced a similar dip around trigger; however, unlike KN, they do not rebound back to higher anomaly scores. This dip is related to the inherent nature of the trigger of a light curve, which often marks the first \textit{real} observation of the transient phase of a light curve, and serves as a reset for the anomaly score. A more detailed analysis of this phenomenon is provided in Appendix \ref{sec:knedip}.

These preliminary findings suggest the potential for enabling real-time identification of anomalous transients. While some known rare classes can be difficult to distinguish from the common classes without a significant amount of data, others can be detected surprisingly soon after trigger. The ability to flag unusual events early in their evolution could prove invaluable for optimizing the allocation of follow-up resources and maximizing the scientific returns from rare transient discoveries.

\subsection{Comparison Against Other Approaches}
\label{sec:benchmarking}

    
\begin{table*}
  
  \hskip-3.3cm
  \begin{center}
  \scalebox{0.76}{
  \begin{tabular}{|l||cccc||ccccc||ccccc|} 
    \hline 
    & \multicolumn{4}{c||}{ \textbf{Transient}}&
    \multicolumn{5}{c||}{ \textbf{Stochastic}}&
    \multicolumn{5}{c|}{ \textbf{Periodic}} \\

    Method  & SLSN & SNII & SNIa & SNIbc & AGN & Blazar & CV/Nova & QSO & YSO & CEP & DSCT & E & RRL & LPV \\ \hline
    
    IForest  & $0.640$ & $0.721$ & $0.428$ & $0.490$ & $0.573$ & $0.710$ & $\mathbf{0.975}$ & $0.468$ & $\mathbf{0.913}$ & $0.359$ & $0.295$ & $0.469$ & $0.549$ & $\mathbf{0.971}$ \\
    \citep{isolationforest}
    & $\pm 0.014$ & $\pm 0.021$ & $\pm 0.032$ & $\pm 0.038$ & $\pm 0.017$ & $\pm 0.009$ & $\mathbf{\pm 0.001}$ & $\pm 0.016$ & $\mathbf{\pm 0.003}$ & $\pm 0.007$ & $\pm 0.012$ & $\pm 0.021$ & $\pm 0.033$ & $\mathbf{\pm 0.007}$ \\ \hline
  
    OCSVM  & $0.577$ & $0.587$ & $0.434$ & $0.492$ & $0.532$ & $0.443$ & $0.909$ & $\mathbf{0.517}$ & $0.792$ & $0.432$ & $\mathbf{0.557}$ & $0.555$ & $0.539$ & $0.943$ \\ \citep{OneClassDef}
    & $\pm 0.014$ & $\pm 0.014$ & $\pm 0.021$ & $\pm 0.011$ & $\pm 0.008$ & $\pm 0.002$ & $\pm 0.001$ & $\mathbf{\pm 0.005}$ & $\pm 0.005$ & $\pm 0.004$ & $\mathbf{\pm 0.005}$ & $\pm 0.003$ & $\pm 0.004$ & $\pm 0.001$ \\ \hline
    
    AE  & $\mathbf{0.736}$ & $\mathbf{0.807}$ & $0.438$ & $0.537$ & $\mathbf{0.701}$ & $\mathbf{0.762}$ & $\mathbf{0.980}$ & $0.443$ & $\mathbf{0.990}$ & $0.564$ & $0.367$ & $\mathbf{0.864}$ & $\mathbf{0.907}$ & $\mathbf{0.996}$ \\
    \citep{rumelhart_1987}
     & $\mathbf{\pm 0.022}$ & $\mathbf{\pm 0.021}$ & $\pm 0.015$ & $\pm 0.019$ & $\mathbf{\pm 0.010}$ & $\mathbf{\pm 0.006}$ & $\mathbf{\pm 0.016}$ & $\pm 0.004$ & $\mathbf{\pm 0.001}$ & $\pm 0.024$ & $\pm 0.015$ & $\mathbf{\pm 0.009}$ & $\mathbf{\pm 0.015}$ & $\mathbf{\pm 0.000}$ \\ \hline
    
    VAE  & $0.669$ & $0.690$ & $0.404$ & $0.522$ & $0.596$ & $0.597$ & $0.849$ & $\mathbf{0.500}$ & $0.795$ & $0.442$ & $0.417$ & $0.561$ & $0.451$ & $0.936$ \\
    \citep{kingma_2013}
     & $\pm 0.015$ & $\pm 0.023$ & $\pm 0.018$ & $\pm 0.025$ & $\pm 0.007$ & $\pm 0.010$ & $\pm 0.028$ & $\mathbf{\pm 0.009}$ & $\pm 0.009$ & $\pm 0.010$ & $\pm 0.007$ & $\pm 0.007$ & $\pm 0.006$ & $\pm 0.007$ \\ \hline
    
    Deep SVDD  & $0.644$ & $0.731$ & $0.475$ & $0.507$ & $0.496$ & $0.607$ & $0.932$ & $0.411$ & $0.901$ & $0.707$ & $0.482$ & $0.636$ & $0.774$ & $0.785$ \\
    \citep{ruff_2018_OneClass}
    & $\pm 0.043$ & $\pm 0.043$ & $\pm 0.040$ & $\pm 0.040$ & $\pm 0.025$ & $\pm 0.044$ & $\pm 0.015$ & $\pm 0.008$ & $\pm 0.022$ & $\pm 0.027$ & $\pm 0.054$ & $\pm 0.055$ & $\pm 0.068$ & $\pm 0.025$ \\ \hline
    
    MCDSVDD  & $\mathbf{0.686}$ & $\mathbf{0.828}$ & $\mathbf{0.624}$ & $\mathbf{0.584}$ & $\mathbf{0.706}$ & $0.512$ & $0.770$ & $0.483$ & $0.854$ & $\mathbf{0.858}$ & $\mathbf{0.819}$ & $\mathbf{0.945}$ & $\mathbf{0.953}$ & $0.953$ \\
    \citep{Perez-Carrasco_2023}
    & $\mathbf{\pm 0.051}$ & $\mathbf{\pm 0.024}$ & $\mathbf{\pm 0.039}$ & $\mathbf{\pm 0.032}$ & $\mathbf{\pm 0.069}$ & $\pm 0.113$ & $\pm 0.127$ & $\pm 0.080$ & $\pm 0.041$ & $\mathbf{\pm 0.025}$ & $\mathbf{\pm 0.015}$ & $\mathbf{\pm 0.006}$ & $\mathbf{\pm 0.003}$ & $\pm 0.008$ \\ \hline

    Classifier + IForest & $\mathbf{0.757}$ & $\mathbf{0.811}$ & $\mathbf{0.619}$ & $0.556$ & $\mathbf{0.715}$ & $\mathbf{0.720}$ & $0.945$ & $0.456$ & $\mathbf{0.977}$ & $\mathbf{0.766}$ & $0.504$ & $\mathbf{0.811}$ & $\mathbf{0.907}$ & $\mathbf{0.969}$ \\
(This work) & $\mathbf{\pm0.047}$ & $\mathbf{\pm0.017}$ & $\mathbf{\pm0.073}$ & $\pm0.039$ & $\mathbf{\pm0.028}$ & $\mathbf{\pm0.032}$ & $\pm0.015$ & $\pm0.041$ & $\mathbf{\pm0.003}$ & $\mathbf{\pm0.066}$ & $\pm0.111$ & $\mathbf{\pm0.038}$ & $\mathbf{\pm0.026}$ & $\mathbf{\pm0.016}$ \\ \hline


    Classifier + MCIF & $0.567$ & $0.699$ & $\mathbf{0.536}$ & $\mathbf{0.560}$ & $0.615$ & $0.701$ & $0.882$ & $\mathbf{0.605}$ & $0.893$ & $\mathbf{0.875}$ & $\mathbf{0.742}$ & $0.773$ & $0.808$ & $0.779$ \\
    (This work) & $\pm0.091$ & $\pm0.046$ & $\mathbf{\pm0.061}$ & $\mathbf{\pm0.034}$ & $\pm0.048$ & $\pm0.045$ & $\pm0.050$ & $\mathbf{\pm0.051}$ & $\pm0.025$ & $\mathbf{\pm0.036}$ & $\mathbf{\pm0.044}$ & $\pm0.031$ & $\pm0.046$ & $\pm0.107$ \\ \hline

    MCIF & $0.503$ & $0.668$ & $0.532$ & $\mathbf{0.643}$ & $0.614$ & $\mathbf{0.745}$ & $\mathbf{0.966}$ & $0.446$ & $0.907$ & $0.514$ & $0.433$ & $0.476$ & $0.447$ & $0.959$ \\  
    (This work) & $\pm0.018$ & $\pm0.008$ & $\pm0.007$ & $\mathbf{\pm0.005}$ & $\pm0.02$ & $\mathbf{\pm0.008}$ & $\mathbf{\pm0.003}$ & $\pm0.007$ & $\pm0.007$ & $\pm0.013$ & $\pm0.009$ & $\pm0.021$ & $\pm0.011$ & $\pm0.004$ \\ \hline

  \end{tabular} }
  \end{center}
  
    \caption{Performance of each model when applied to the dataset used in \citet{Perez-Carrasco_2023}. Each row represents a different anomaly detection algorithm and each column represents a different class being chosen as the anomalous class. The performance is evaluated using the AUROC score of detected anomalies. The top 3 metrics per class are marked in bold. The AUROC scores for the first 5 methods are taken directly from and are reported in \citet{Perez-Carrasco_2023}. A visual representation of this table is shown in Figure \ref{fig:visual_table}. }
  \label{table:results}
\end{table*}

In the field of anomaly detection in time-domain astronomy, there is no comprehensive baseline on which to evaluate different detection methods. This is largely because of the vastly differing definitions of what \textit{anomaly detection} is, for example, the difference between unsupervised and novelty detection methods as described in Section \ref{sec:introduction}. Baselining all existing anomaly detection methods is a much-needed line of future work, especially as there is no consensus on which method will work best on the deluge of data that will be available when LSST is running.

Despite these challenges, \citet{Perez-Carrasco_2023} evaluated 5 different approaches to anomaly detection (see Table \ref{table:results} for all benchmarked approaches), and we use their dataset (which was inspired by \citealt{SanchezSaez2021}) to benchmark our classifier-based approach. In contrast to our dataset of raw light curve data, this dataset consists of tabular \textit{features} extracted from light curves. We evaluate three new techniques for anomaly detection on this dataset: using a classifier with \texttt{MCIF}, a classifier with just a single isolation forest, and \texttt{MCIF} on its own\footnote{We can use \texttt{MCIF} on its own as this is a dataset of features extracted from time-series, not the raw time-series.}. The dataset is split into 3 hierarchical categories with 4-5 transient classes each. Evaluation is performed separately for each class, each time counting that transient class as anomalous and the rest of its hierarchical category as common. Full evaluation is performed across 5 folds of testing data for cross-validation.

As seen in Table \ref{table:results} (and visually in Figure \ref{fig:visual_table}), our classifier-based approach with an isolation forest is one of the top approaches for most transient classes, showing the power of using a classifier's latent space for anomaly detection. Using a classifier with \texttt{MCIF} also preforms promisingly, however is sometimes worse than using a classifier with a single isolation forest. This is not the case on our dataset and is discussed further in Appendix \ref{sec:MCIF_Advantages}.


\subsection{Scaling the Latent Space}
\label{sec:hyperparameter}


Anomaly detection poses a unique challenge for model evaluation due to the nature of unsupervised learning: true anomalies are only revealed during the final testing phase. Consequently, we refrain from tuning hyperparameters for model selection and instead retrospectively analyze the effects of different hyperparameter choices, particularly the size of the latent space without risking overfitting during the development process.

\begin{figure}
\centering
\includegraphics[width=0.45\textwidth]{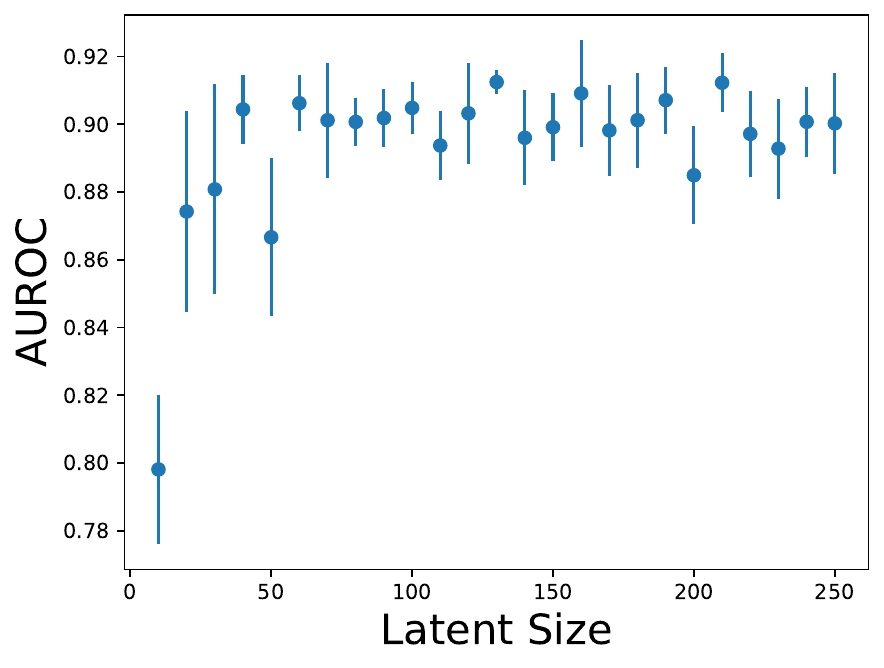}
\caption{Anomaly detection performance (AUROC) of models trained with different latent space sizes. A significant improvement is observed when increasing the latent size up to 50 dimensions, with performance plateauing thereafter.}
\label{fig:scaling}
\end{figure}

To assess the impact of latent space size on anomaly detection performance, we train multiple models with varying latent dimensions and evaluate them using the AUROC. As shown in Figure \ref{fig:scaling}, increasing the latent size beyond 50 leads to significant improvements in anomaly detection performance, with diminishing returns thereafter. Smaller models generally exhibit lower average performance and higher variance. Interestingly, we do not observe a performance drop in high-dimensional latent spaces, despite the presence of numerous correlated features. This robustness can be attributed to the effectiveness of isolation forests and tree-based algorithms in handling high-dimensional data by selectively identifying the most discriminative features for data partitioning. 

Based on Figure \ref{fig:scaling}, we select a latent size of 100 neurons for this work, as it yields a high anomaly detection AUROC score. However, it's worth noting that this choice is somewhat arbitrary, as any sufficiently large latent size beyond 50 dimensions appears to provide comparable performance.

Interestingly, while classifiers prove effective for anomaly detection, we do not find a significant correlation between classification accuracy and anomaly detection performance. This highlights a key interpretability challenge in both traditional neural networks and our approach, warranting further investigation.


\section{Conclusion}
\label{sec:Conclusion}
The advent of large-scale wide-field surveys has revolutionized time-domain astronomy, unveiling an extraordinary diversity of astrophysical phenomena. Surveys such as the Zwicky Transient Facility and the upcoming Legacy Survey of Space and Time conducted by the Vera C. Rubin Observatory promise to discover transients in vast numbers, with LSST expected to generate millions of alerts each night. This deluge of data presents both a challenge and an opportunity: while the sheer volume of detections makes manual inspection infeasible, it also offers the potential to discover entirely new classes of transients that are rare or have remained hidden in previous surveys. To fully harness the potential of these wide-field surveys, automated methods for real-time anomaly detection are essential. By identifying and prioritizing the most unusual and interesting events amid the flood of data, these techniques can facilitate rapid follow-up and characterization, enabling us to deepen our understanding of the time-domain universe.

The most common approach for transient anomaly detection is to construct a feature extractor that can map light curves to a low-dimensional latent space, and then apply clustering or outlier detection algorithms to identify anomalies within that space. Constructing a latent space that represents transients well for anomaly detection is difficult, and most previous approaches have either user-defined features or unsupervised deep-learning approaches.

In this work, we have introduced a novel approach that leverages the latent space of a neural network classifier for identifying anomalous transients. Our pipeline, which combines a deep recurrent neural network classifier with our novel Multi-Class Isolation Forest (\texttt{MCIF}) anomaly detection method, demonstrates promising performance on simulated data matched to the characteristics of the Zwicky Transient Facility.

The key advantages of our approach are:
\begin{enumerate}

\item The RNN classifier maps light curves into a low-dimensional latent space that naturally clusters similar transient classes together, providing an effective representation for anomaly detection. We repurposed the penultimate layer of this classifier as the feature space for anomaly detection.

\item Our novel \texttt{MCIF} method addresses the limitations of using a single isolation forest on the complex latent space by training separate isolation forests for each known transient class and taking the minimum score as the final anomaly score.

\item Our classifier input format eliminates the need for interpolation by incorporating time and passband information, enabling the model to learn inter-passband relationships and handle irregular sampling.

\end{enumerate}

To mimic a real-world scenario, we evaluated our approach on a realistic simulated dataset containing 12,040 common transients and 54 anomalous events. After following up \texttt{MCIF}'s top 2000 ranked transients, we accurately identified $41 \pm 3$ out of the 54 true anomalies. That is, after following up the top 15\% highest ranked scores, we recovered 75\% of the true anomalies. CaRTs look very similar to common supernovae, and thus are difficult to identify. If we exclude CaRTs from our anomalous sample, our recovery of anomalies increases sharply to $87\%$ ($47 \pm 2$ out of the 54 true anomalies) after following up the top 2000 ($\sim 15\%$) highest-scoring transients.

The learned latent space exhibits clear separation between common and anomalous transient classes, and our preliminary analysis suggests the potential for real-time anomaly detection using limited early-time observations. The pre-trigger information encoded by our RNN enables our model to identify anomalous transients at early stages in the light curve, and even by trigger, a significant separation between common and anomalous transients is captured. In particular, KN, PISN, and ILOT all stand out as anomalous shortly after the time of trigger.

Future work encompasses several promising directions. Firstly, benchmarking our model against other similar approaches is important for a comprehensive performance assessment. Currently, comparing models is difficult, because a standard test dataset of anomalies does not exist. Developing a realistic benchmark dataset that encompasses a representative population of common and example anomalous transients will improve the quality of methods developed by the community and enable robust evaluation metrics. Moreover, a detailed comparative analysis of \texttt{MCIF} with previous class-by-class anomaly detection approaches should be carried out to gain a deeper understanding of their relative strengths and limitations in this domain. 

Secondly, integrating techniques from other anomaly detection methods, such as active learning \citep{Lochner2020Astronomaly}, could help to distinguish new anomalies as \textit{interesting} or not. Beyond direct anomaly detection, \texttt{MCIF} can be used to identify which known class an anomalous object most closely resembles based on the individual isolation forest scores. Additionally, we plan to apply the proposed architecture to real observational data, moving beyond simulations and testing the model's effectiveness in a practical astronomical context.

A significant contribution of this work is the demonstration that a well-trained classifier can be effectively repurposed for anomaly detection by leveraging the clustering properties of its latent space. The flexibility of our approach allows for the adaptation of any classifier to an anomaly detector. For example, using existing classifiers as feature extractors for spectra, images, or time series from other domains, we can build effective anomaly detectors.

Another significant advantage of our approach is that the clustering properties of the latent space extend to unseen data, enabling few-shot classification of astronomical transients with limited labeled examples. This will be useful for the early observations from new surveys such as LSST. Furthermore, our input method lends itself well to transfer learning from one survey to another because it explicitly uses the passband wavelength. Future work should explore transfer learning from ZTF data to other surveys such as PanSTARRS or LSST simulations. 

In conclusion, our novel approach to real-time anomaly detection in astronomical light curves, combining a deep neural network classifier with Multi-Class Isolation Forests, demonstrates the power of leveraging well-clustered latent space representations for identifying rare and unusual transients. As the era of large-scale astronomical surveys continues to produce unprecedented volumes of data, the development and refinement of such techniques will be crucial for making discoveries in time-domain astronomy.

\section*{Acknowledgements}

We would like to thank the Cambridge Centre for International Research (CCIR) for fostering this collaboration. ML acknowledges support from the South African Radio Astronomy Observatory and the National Research Foundation (NRF) towards this research. Opinions expressed and conclusions arrived at, are those of the authors and are not necessarily to be attributed to the NRF.  

This work made use of the \texttt{python} programming language and the following packages: \texttt{numpy} \citep{numpy}, \texttt{matplotlib} \citep{matplotlib}, \texttt{seaborn} \citep{seaborn}, \texttt{scikit-learn} \citep{scikit-learn}, \texttt{pandas} \citep{pandas}, \texttt{astropy} \citep{astropy}, \texttt{umap-learn} \citep{umap-learn}, \texttt{keras} \citep{keras}, and \texttt{tensorflow} \citep{tensorflow}.

We acknowledge the use of the ilifu cloud computing facility – \href{www.ilifu.ac.za}{www.ilifu.ac.za}, a partnership between the University of Cape Town, the University of the Western Cape, Stellenbosch University, Sol Plaatje University and the Cape Peninsula University of Technology. The ilifu facility is supported by contributions from the Inter-University Institute for Data Intensive Astronomy (IDIA – a partnership between the University of Cape Town, the University of Pretoria and the University of the Western Cape), the Computational Biology division at UCT and the Data Intensive Research Initiative of South Africa (DIRISA).

This work used Bridges-2 at Pittsburgh Supercomputing Center through allocation PHY240105 from the Advanced Cyberinfrastructure Coordination Ecosystem: Services \& Support (ACCESS) program \citep{NSF-ACCESS-Boerner2023}, which is supported by U.S. National Science Foundation grants \#2138259, \#2138286, \#2138307, \#2137603, and \#2138296.
\section*{Data Availability}

The code used in this work is \href{https://github.com/Rithwik-G/AstroMCAD}{publicly available}. The models used to create the simulations that generate the data used in this work were released in PLAsTiCC \citep{KesslerPlasticcModels} and are available at \href{https://zenodo.org/record/2612896\#.YYAz1NbMJhE}{https://zenodo.org/record/2612896\#.YYAz1NbMJhE}. To generate light curves following ZTF observing properties, we use the \texttt{SNANA} software package \citep{SNANA}, developed for PLAsTiCC, with observing logs from the ZTF survey to generate data. A version of these simulations was first used in \citep{Muthukrishna19RAPID} and have since been updated to resolve a known problem with core-collapse SNe. The data is publicly available upon reasonable request to the corresponding author. 



\bibliography{paper}
\bibliographystyle{mnras}



\appendix

\section{Advantages of \texttt{MCIF}}

\begin{figure}
\centering
\includegraphics[width=0.48\textwidth]{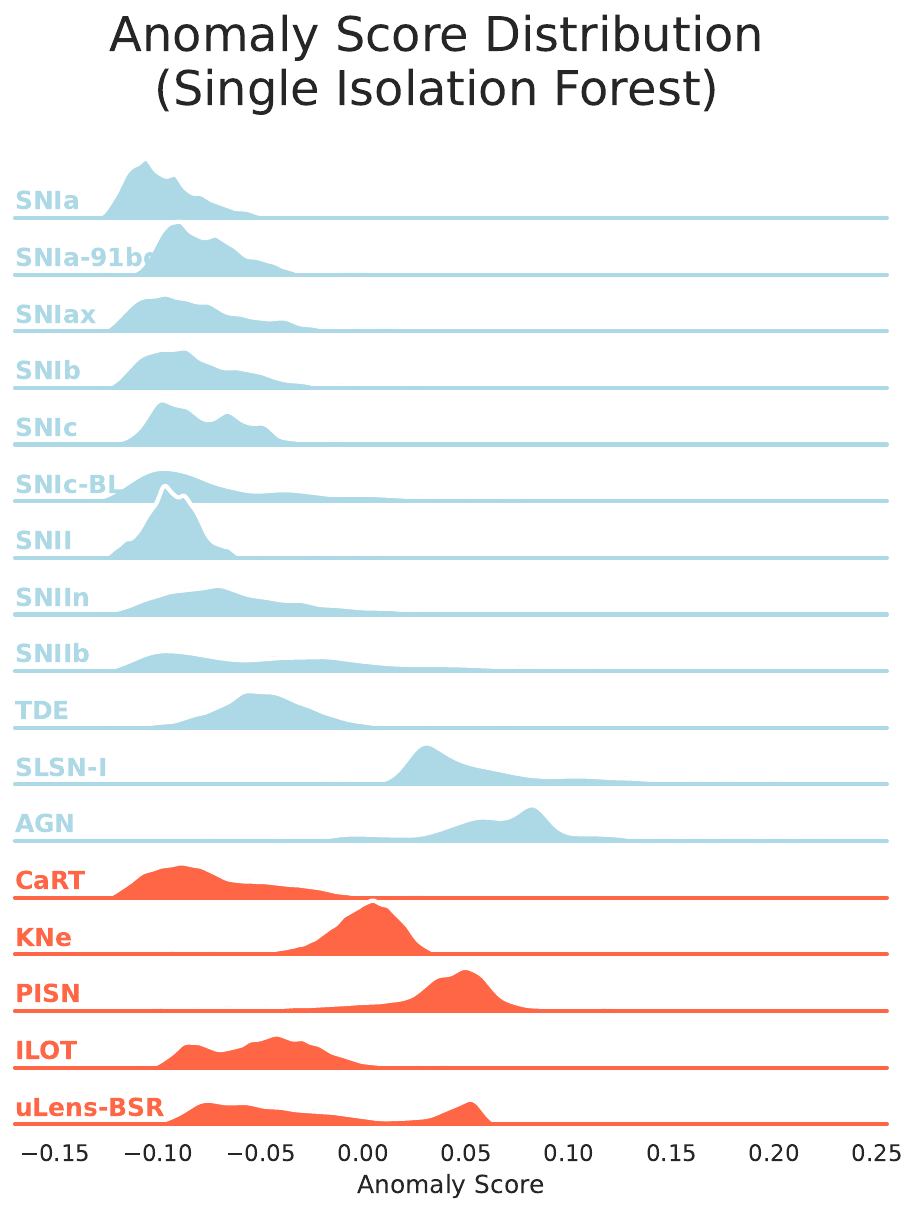}
\caption{The distribution of anomaly scores for full light curves when using a single isolation forest for anomaly detection. The scores are derived from the unseen anomalous data and the common transient testing data. The bottom 5 classes (in red) are the anomalous classes. There is some separation between the anomaly scores of common and anomalous classes, but certain common classes are considered very anomalous (unlike when using \texttt{MCIF} as seen in Figure \ref{fig:Distribution}).}
\label{fig:SingleIsoDistribution}
\end{figure}

\label{sec:MCIF_Advantages}

\begin{figure*}

\centering

\begin{tabular} {@{}c@{}}
\includegraphics[width=0.8\textwidth]{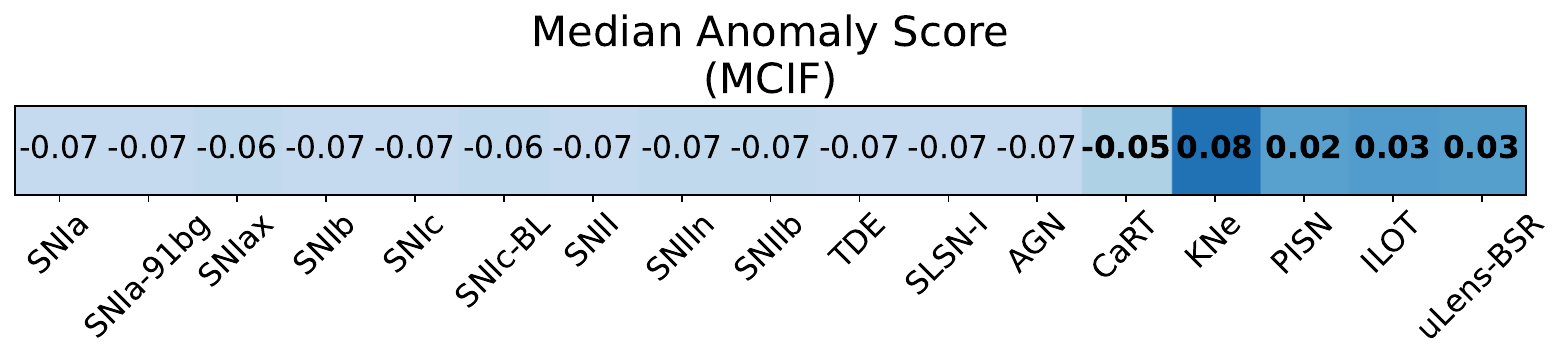}
\end{tabular}

\begin{tabular} {@{}c@{}}
\includegraphics[width=0.8\textwidth]{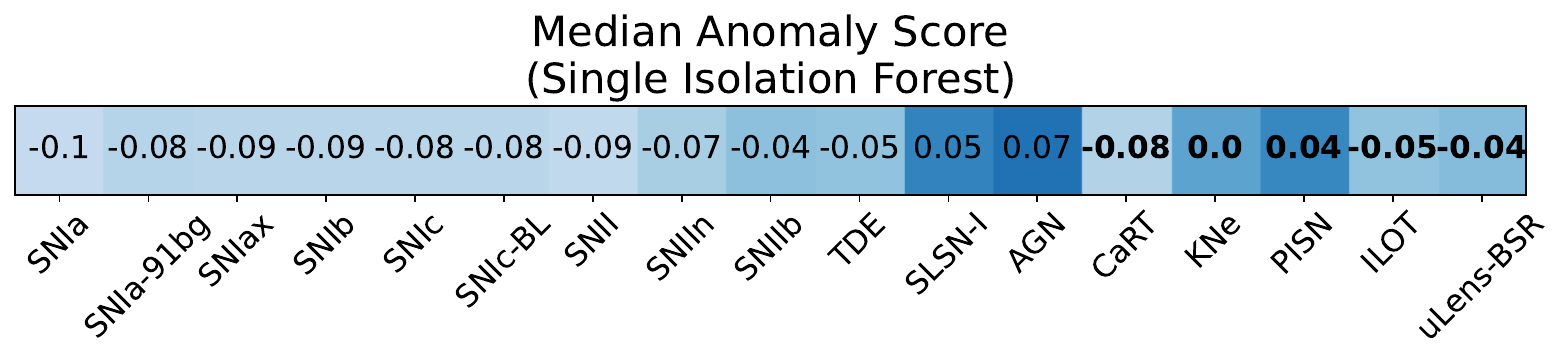}
\end{tabular}

\caption{The median anomaly score for each class computed for latent representations of transients obtained from full light curves when a single isolation forest is used for anomaly detection [bottom] and when \texttt{MCIF} is used [top] (this is the exact same as Figure \ref{fig:MCIFAverageScore}, reproduced for convenience). The scores are derived from the unseen anomalous data and the common transient testing data. The 5 classes on the right (scores in bold) are anomalous. The common classes have somewhat lower median scores when using a single isolation forest, but the common classes SLSN-I and AGN (among others) are considered very anomalous, unlike when using \texttt{MCIF}.}
\label{fig:FullAverageScore}
\end{figure*}

\begin{figure*}
\centering
\begin{tabular}{ll}
\includegraphics[width=0.45\textwidth]{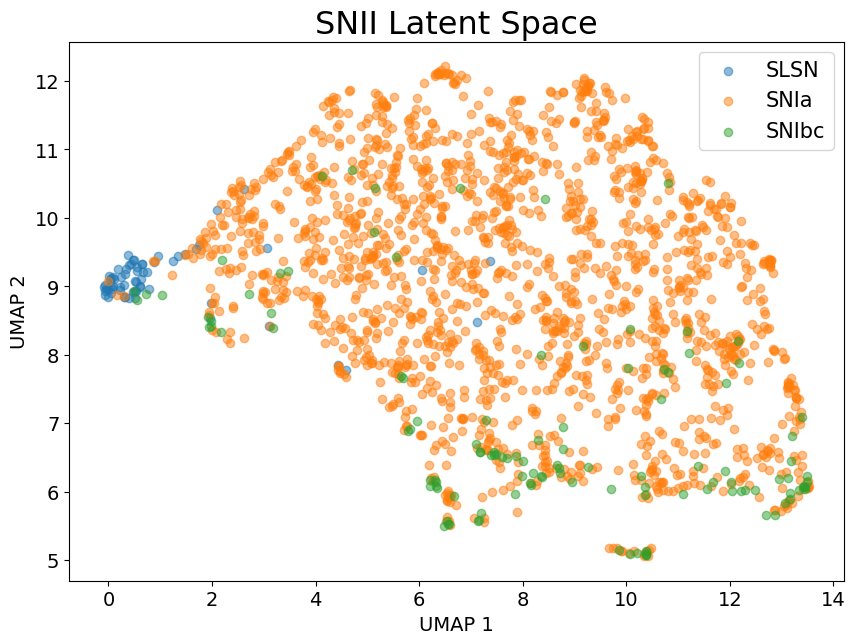}
&
\includegraphics[width=0.45\textwidth]{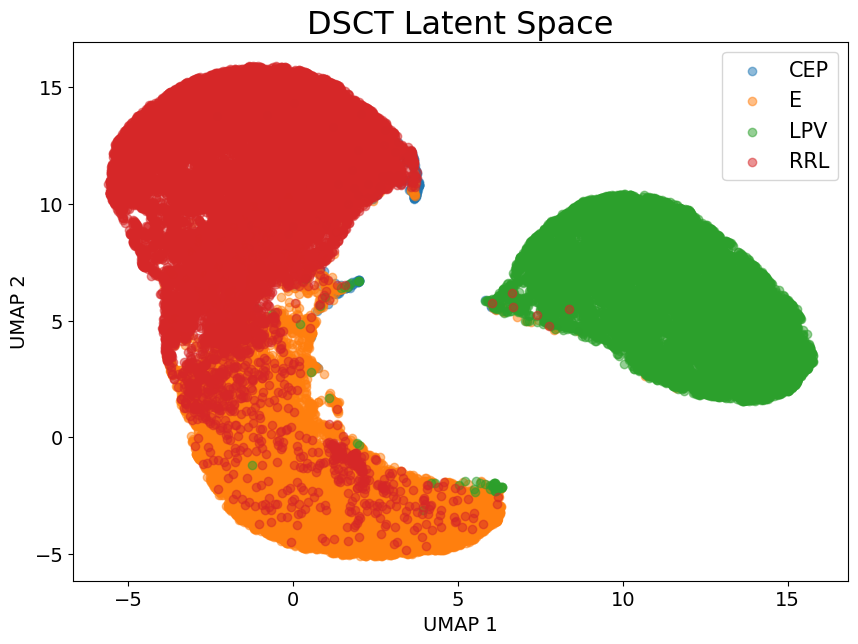}
\end{tabular}
\caption{The UMAP reduction of the training data in the latent space for a classifier trained for detecting the class SNII [left] and DSCT [right] as anomalous using the data introduced in \citealp{Perez-Carrasco_2023} and used in Section \ref{sec:benchmarking}. As the UMAP only plots the training data, it includes all the classes in the respective hierarchical category (seen in Table \ref{table:results}) but the one set aside as anomalous.}
\label{fig:umapres}
\end{figure*}

Before proposing the \texttt{MCIF} pipeline, we attempted to use a normal isolation forest to detect anomalies from the latent representation $z_s$ of a light curve. We trained an isolation forest on all the common classes of our training data using $200$ estimators. To account for the class imbalance in our training data, we weighted samples from underrepresented classes more heavily during the training of the isolation forest, using the same weighting scheme as in Equation \ref{eq:1}. The anomaly score function $A(Z_s)$ was simply the negated anomaly score output from a single isolation forest trained on all the latent representations of the training data.

As shown in Figure \ref{fig:FullAverageScore}, there is little distinction in the anomaly scores of most anomalous and common classes when using a single isolation forest. Surprisingly, the common classes SLSN-I and AGN are classified as relatively more anomalous than all the other classes. The distribution of anomaly scores in Figure \ref{fig:SingleIsoDistribution} reveals that although there is overall separation between common and anomalous classes, certain common classes are classified as very anomalous.

The UMAP reduction of the latent space, as depicted in Figure \ref{fig:UMAP}, provides insight into this behaviour. The SLSN-I and AGN classes are significantly distant from the main cluster formed by other classes. This isolation from the central cluster may explain the high anomaly scores associated with these classes. This hypothesis is also supported by the near-perfect classification of these classes, shown in the confusion matrix and ROC curves in Figure \ref{fig:ConfusionROC}. In fact, the near-perfect classification hinted towards this poor result in anomaly detection, showing us that these transients are easy to separate from other classes, and hence are also easy to mark as anomalous. In summary, while an isolation forest is good at detecting anomalies, it struggles to capture the structure of a latent space with numerous clusters. This drawback of using a single isolation forest could explain why other works report high anomaly scores for SLSN-I and AGN \citep[e.g.][]{vraenn}. Using a class-by-class (or cluster-by-cluster) anomaly detector, such as \texttt{MCIF}, can mitigate this. Directly comparing Figure \ref{fig:Distribution} and Figure \ref{fig:SingleIsoDistribution} empirically demonstrates the advantages of \texttt{MCIF}.

Further analysis of MCIF's performance on the comparative evaluation dataset (Section \ref{sec:benchmarking}) reveals that, contrary to the results shown in Figure \ref{fig:Distribution}, a single isolation forest generally outperforms MCIF (Table \ref{table:results}). Investigating the UMAP representations of the latent space for classes exhibiting this discrepancy offers insights. When SNII is considered anomalous, the latent space (Figure \ref{fig:umapres} [left]) lacks clear separation between SNIbc and SNIa, likely due to poor generalization caused by the limited number of SNIbc transients in the training set, explaining the single isolation forest's superior performance. However, for the DSCT class (Figure \ref{fig:umapres} [right]), distinct visual clusters are present, and MCIF achieves better results. These findings suggest that MCIF enhances performance when majority classes are well-separated, a characteristic seemingly inherent to the dataset rather than the classifier-based latent space identification approach, as a single isolation forest surpasses MCIF on the raw data for most classes where it also outperforms MCIF on the classifier's latent space. Future research should explore the factors influencing MCIF's effectiveness based on the separability of raw data, with the SNII case indicating a partial dependence on data quantity, as increased data improves the DNN's generalization ability.

\section{Visual Comparison to other Approaches}

\begin{figure*}
    \centering

    \includegraphics[width=\textwidth]{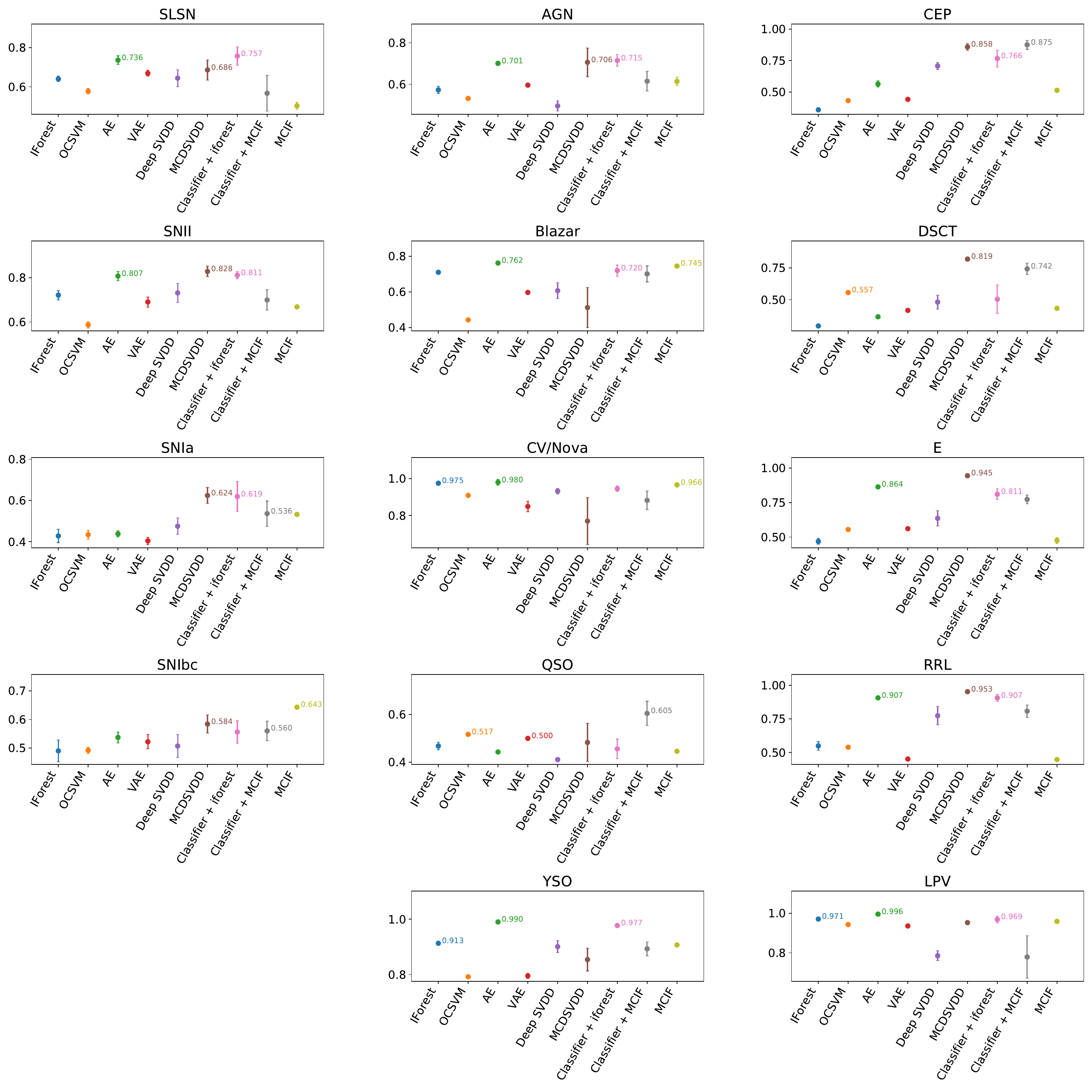}

    \caption{Visual representation of the comparative analysis depicted in Table \ref{table:results}. The AUROC is written for the models top 3 models for each class.}
    \label{fig:visual_table}
\end{figure*}

Figure \ref{fig:visual_table} is a visual representation of the results depicted in Table \ref{table:results}.

\section{The Kilonova Dip}

\label{sec:knedip}

\begin{figure*}
\centering
  \begin{tabular}{lll}
  \includegraphics[width=0.3\textwidth]{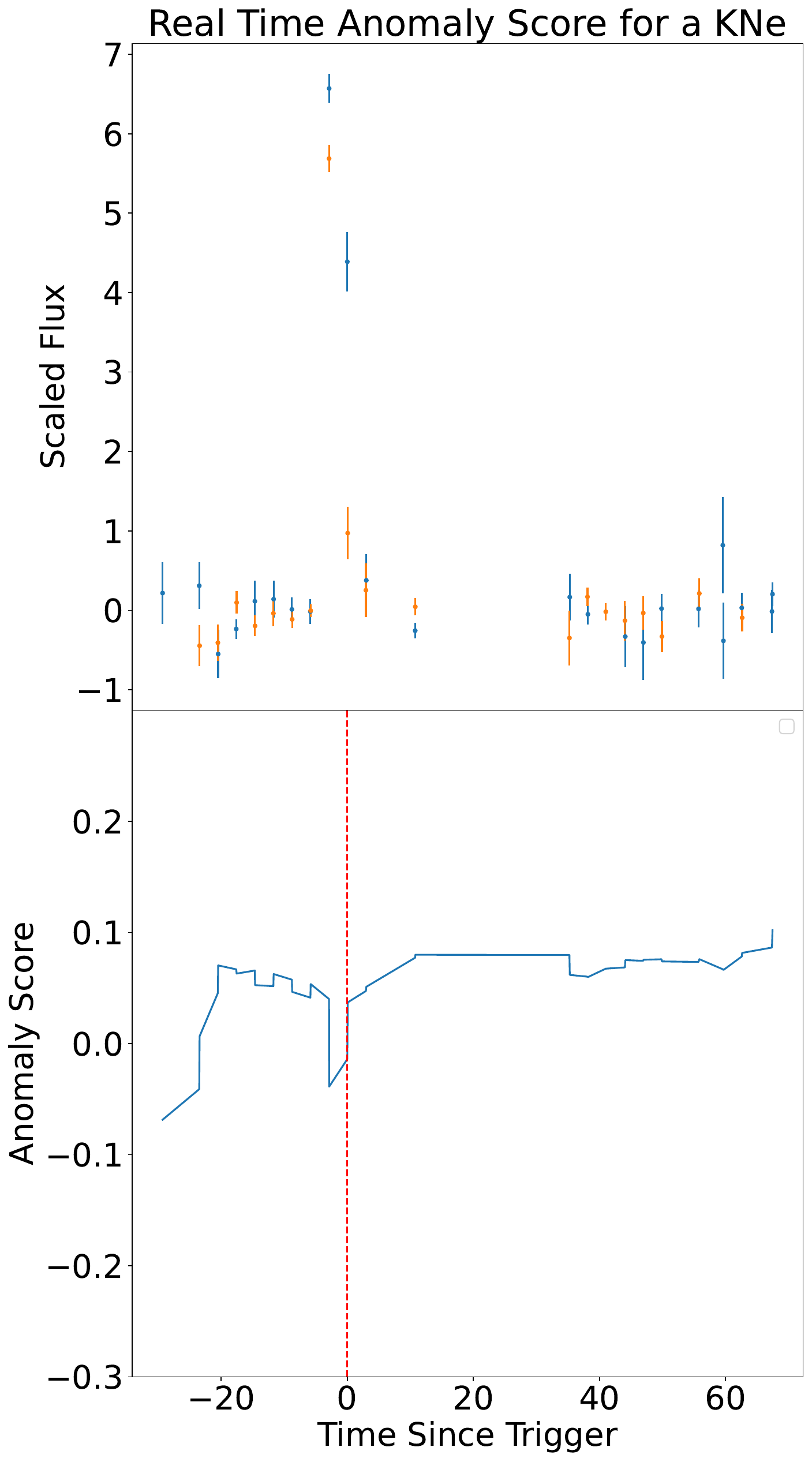}  
  &
  \includegraphics[width=0.3155\textwidth]{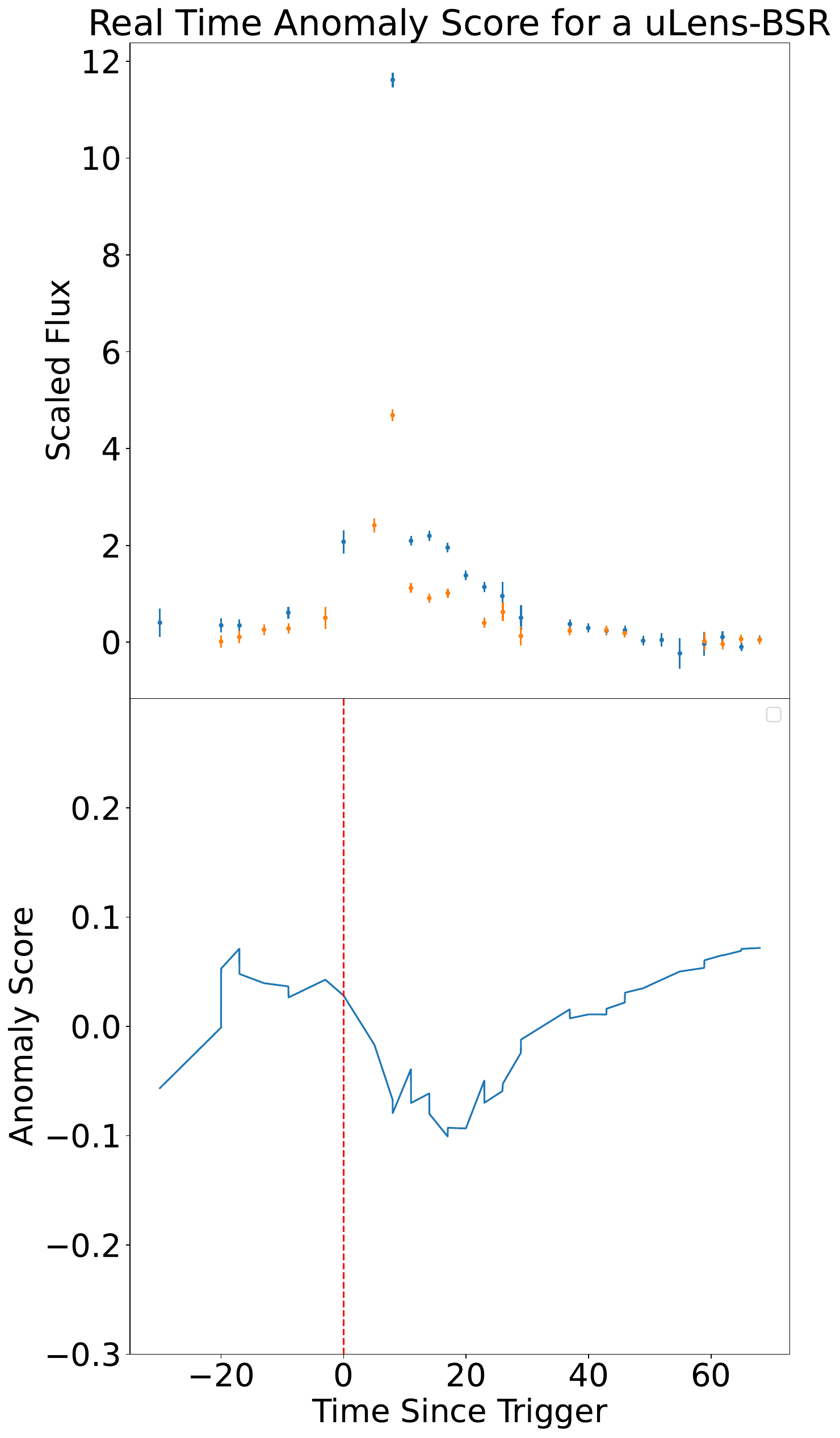}
  &
  \includegraphics[width=0.3\textwidth]{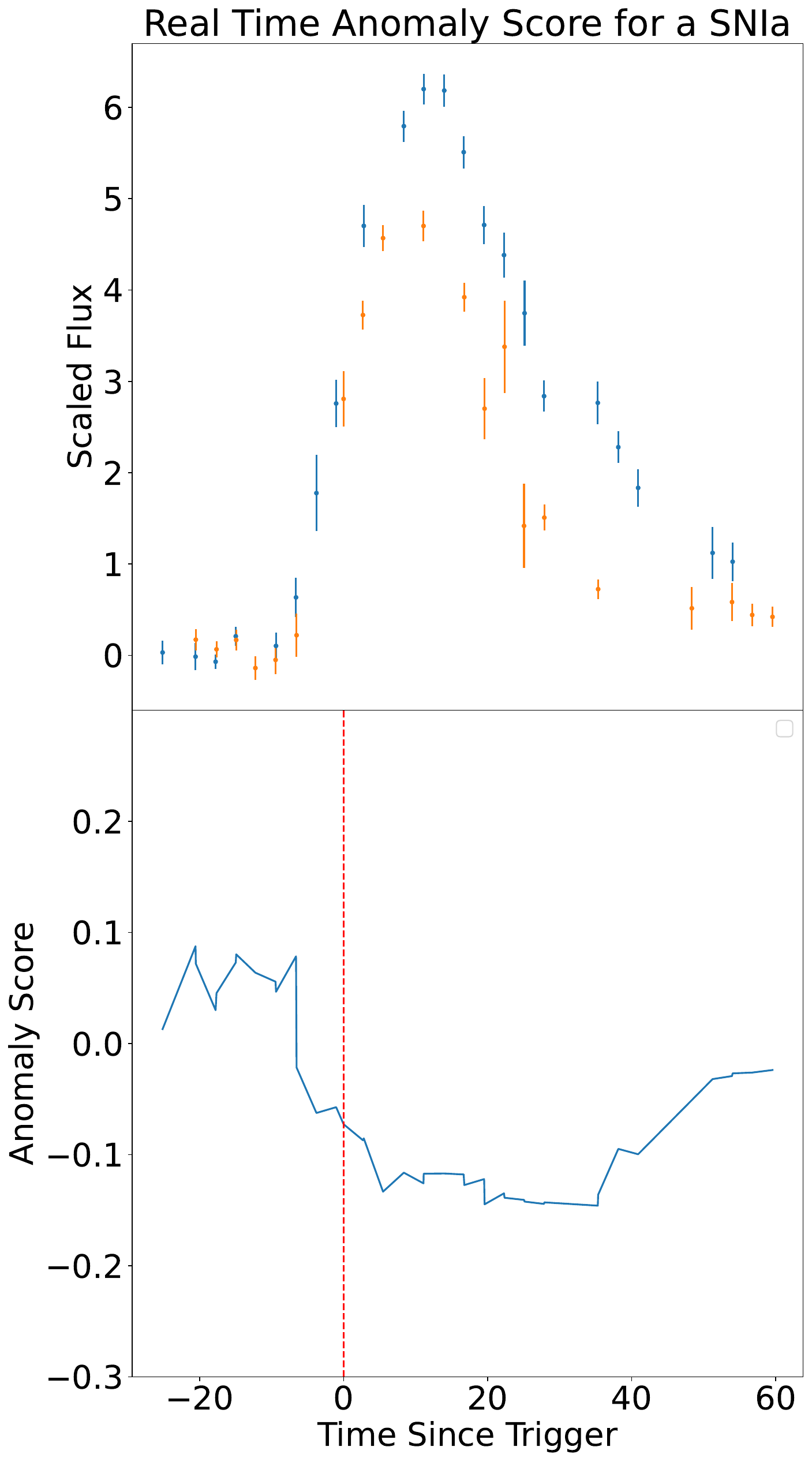}  

    \end{tabular}
    \caption{Real-time anomaly scores for a sample KN, uLens-BSR, and SNIa. They all exhibit a significant dip near trigger, but the dip for the SNIa is followed by a further decline, whereas KN and uLens-BSR show a sharp increase after the dip. In light curves in the top panels, the blue markers represent the g-band and the orange markers represent the r-band fluxes.}

    \label{fig:anomexamples}

\end{figure*}

\begin{figure*}
\centering
  \begin{tabular}{ll}
  \includegraphics[width=0.45\textwidth]{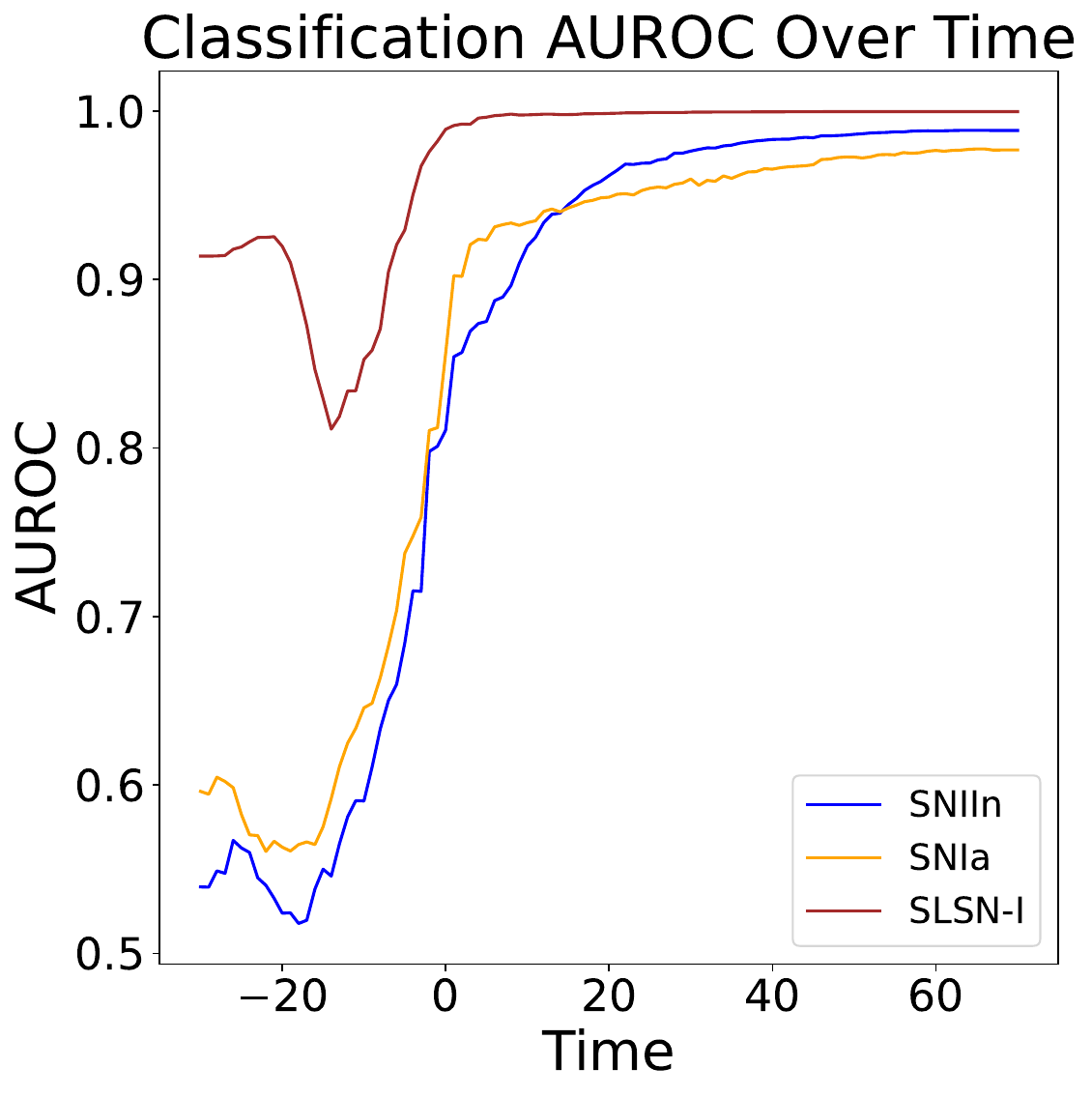}  
  &
  \includegraphics[width=0.5\textwidth]{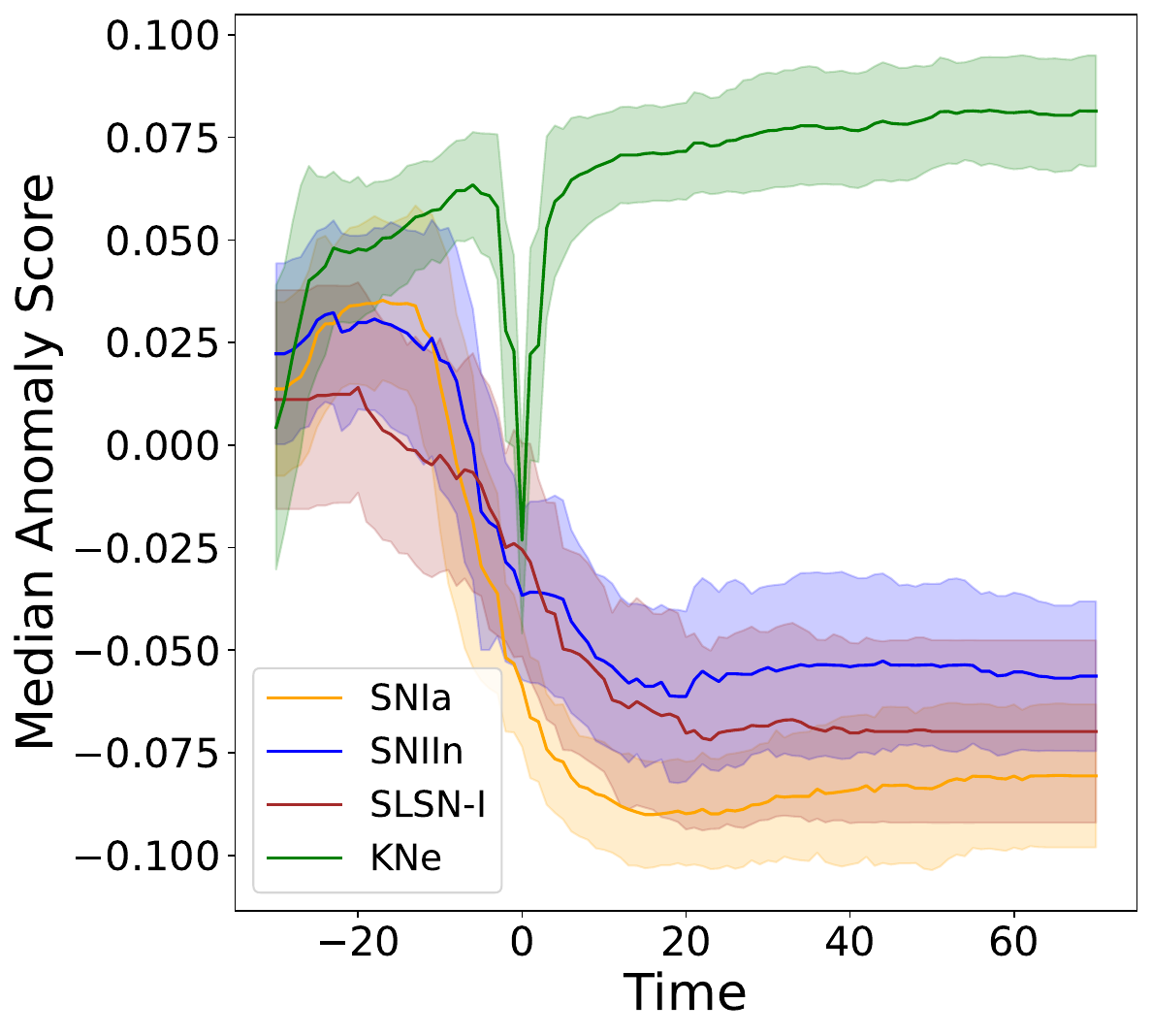}  

    \end{tabular}

    \caption{Real-time AUROC values for selected classes [left] and real-time anomaly scores for a different subset of classes than Figure \ref{fig:RealTimeAnalysis} [right]. Classes that are poorly classified pre-trigger (e.g. SNIIn, and SNIa) exhibit a dip in anomaly score similar to KN, which coincides precisely with the sudden increase in classification accuracy.}
    \label{fig:otherealtime}
\end{figure*}

As illustrated in Figure \ref{fig:RealTimeAnalysis}, there is an unusual dip in the anomaly scores of KN around the trigger time. Further analysis reveals that most common classes also experience a similar dip at trigger, but they do not rebound to high anomaly scores afterward. Examples are shown in Figure \ref{fig:anomexamples}.

The trigger of a light curve often corresponds to the first observation detected as part of the transient phase of the light curve, and very few common classes can be effectively classified before trigger. Effective pre-trigger classification is likely due to host galaxy information (host redshift and Milky Way extinction) or the periodic nature of certain transient events (e.g. AGNs) which means they are midway through their evolution at trigger. Figure \ref{fig:otherealtime}, shows that classes with a high pre-trigger classification accuracy (e.g. SLSN-I) have consistently declining anomaly scores before trigger. In contrast, classes with poor pre-trigger classification (e.g. SNIIn and SNIa) exhibit a slight increase in anomaly scores before trigger, followed by a sudden dip. This sudden dip resembles the behaviour of KN and coincides perfectly with the sudden jump in classification performance. This suggests that our pipeline struggles to detect KN as anomalous before trigger for the same reasons it is unable to classify SNIIn before trigger. Despite KN exhibiting a slight upward trend before trigger, it seems that the new observation near trigger means much more (likely due to the high S/N of that observation).

For observations after trigger, we found that the anomaly score ``resets'' to mark the true beginning of the transient phase. For example, in the case of KN, the high S/N trigger observation signals the actual start of the transient and resets the anomaly score. Subsequent observations, characterised by a sudden decline back to the background level, quickly push KN into the anomalous category (as short-time-scale events are rare). However, in the case of poorly classified common classes, this reset is followed by a further decline in anomaly score as classification accuracy increases, making the dip appear normal. A similar effect can be seen in other anomalous classes, most notably in uLens-BSR transients. The result is less pronounced as the first rise of uLens-BSR transients does not always coincide with trigger, leading to a distributed dip around trigger for uLens-BSR in Figure \ref{fig:RealTimeAnalysis} and a dip offset from trigger in Figure \ref{fig:anomexamples}.

\section{Example Light Curves}
\label{sec:dataset}

\begin{figure*}
    \centering
    \begin{tabular}{ll}
      \includegraphics[scale=0.28]{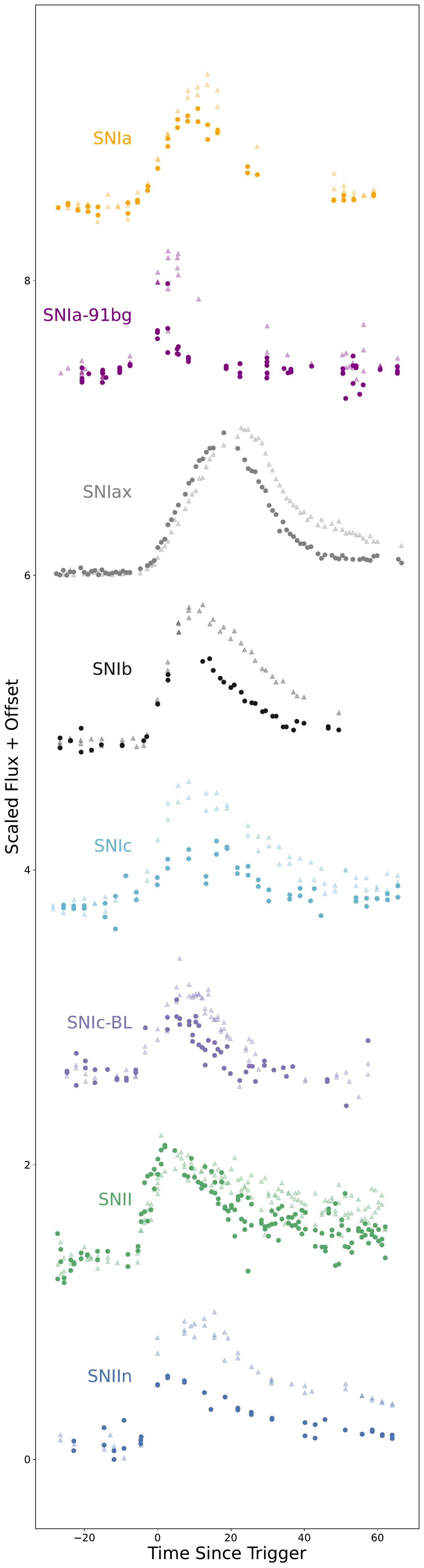}  
      &
      \includegraphics[scale=0.28]{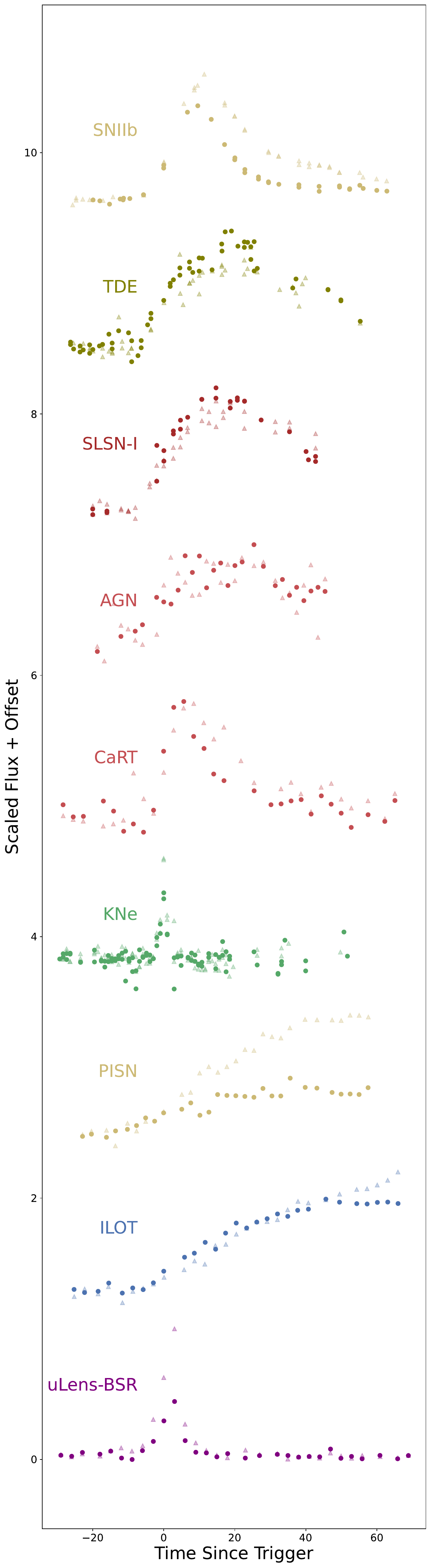}  
    \end{tabular}
    
    \caption{Sample light curves from each transient class used in this work. We only plot transients with low signal-to-noise and low host redshift ($z < 0.5$) to help visually compare shapes. The dark circular markers represent the r band while the light triangular markers represent the g band. Flux errors are not plotted.}
    \label{fig:samplecurves}
\end{figure*}

A sample light curve from each class is illustrated in Figure \ref{fig:samplecurves}.

\bsp	
\label{lastpage}
\end{document}